%% file: appleappstore.tex
\documentclass[10pt,a4paper]{IEEEtran}
\usepackage{graphicx}
\usepackage{amssymb}
\usepackage{epstopdf}
\usepackage[utf8]{inputenc}
\usepackage[T1]{fontenc}

\usepackage{courier}
\usepackage{color}
\pagecolor{white}

\usepackage{fvextra}
\newcommand{\vv}[1]{\path|#1|}

\usepackage{parskip}
\usepackage{xurl}
\usepackage{enumitem}
\usepackage[table]{xcolor}

\usepackage{subcaption}
\usepackage{hyperref}
\usepackage{tablefootnote}

\input{listings}

\title{Poor Privacy Practices Of The Apple App Store: Cookies, Advertising and Tracking Of Users}
\author{D.J.Leith\thanks{Contact email: doug.leith@tcd.ie}\\Trinity College Dublin, Ireland\\March 31st 2026}

\begin{document}
\maketitle
\begin{abstract}
We analyse the data that the Apple App Store sends to and receives from Apple servers.    We find that multiple cookies are sent by Apple servers and stored on the handset.  Adverts with tracking identifiers are also stored on the handset, and we observe that adverts are selected by Apple servers using GDPR special category personal data such as sexuality, religious beliefs and health.   User interactions with the Apple App Store (apps and adverts viewed, buttons clicked, searches made etc) are transmitted to Apple servers alongside identifiers linking this data to the individual user and device.   We show that much of this data storage and transmission is not essential for the service requested by the user, namely to view, search and install apps.  No consent is sought for any of this data storage and processing, and there is no opt out.


\end{abstract}


\section{Introduction}
We analyse the data that the Apple App Store running on an iPhone sends to and receives from Apple servers.  
In summary, we find that:
\begin{enumerate}
\item Numerous cookies are sent by Apple servers and stored on the handset.   We could find no public documentation on these cookies, the handset user is not asked to consent to their storage, there is no opt out, and all but one of the cookies are not essential for the services explicitly requested by the user\footnote{\label{foot:notessential}When we block transmissions the can still view, search, install and open apps in the Apple App Store i.e. perform the primary functions of any app store including the Apple App Store.}

\item Apple servers send adverts which are stored on the handset by the Apple App Store.  The adverts include tracking identifiers that can be linked to the individual user and device.  The handset user is not asked to consent to this storage, and there is no opt out.  Storage of adverts by Apple is self-evidently not essential for the services explicitly requested by the user.

\item When the user searches for apps, the Apple servers use sensitive personal data relating to the user's sexuality, religious beliefs and health, to target adverts.   The user is not asked to consent to this use of their personal data by Apple for advertising, and there is no opt out.   

\item User interactions with the Apple App Store (apps and adverts viewed, buttons clicked, searches made etc) are transmitted to Apple servers.   This tracking of how people use their iPhone is not anonymous, with user and device identifiers sent alongside the data allowing it to be linked to the individual user and device.  The purpose of this tracking is unclear but (i) it is not essential for the service requested by the user\footref{foot:notessential} and (ii) data collected for detecting software bugs etc could be anonymous.   No consent is sought and there is no opt out from this data collection by Apple.
\end{enumerate}

Our measurements were carried out in Ireland, which is in the EU.  We discuss the Irish and EU legal context more in Section \ref{sec:legal_short} below, but we briefly make the following observations here:

\begin{enumerate}
\item \textit{Storage Is Not Essential}.
Storage of cookies and other data on handsets is regulated by the EU e-Privacy Directive.   Consent is required for storage except when ``\emph{strictly necessary in order to provide an information society service explicitly requested by the subscriber or user}''\footnote{See \nolinkurl{https://eur-lex.europa.eu/LexUriServ/LexUriServ.do?uri=CELEX:32002L0058:en:HTML}.  There is also a second exception ``\emph{`for the sole purpose of carrying out the transmission of a communication over an electronic communications network}'', but that is not relevant here.}.   For this strictly necessary exemption to apply: (i) the user must have taken a positive action to request a service, and (ii) the service does not work if the storage is disabled~\cite{edpbcookies2012}.   The storage that we observe does not appear to meet these conditions\footref{foot:notessential}.   Note that legitimate interest is not a valid legal basis for storing cookies and other data, and the EU e-Privacy Directive applies to all data storage not just storage of personal data.

\item \textit{Processing Of GDPR Special Categories Data For Advertising}.
Processing of personal data is regulated by the EU GDPR~\cite{gdpr}.  The data that we observe being stored and transmitted is all personal since user and device identifiers are sent alongside allowing it to be linked to the individual user and device.   The GDPR provides extra protection for so-called \emph{special category} data, which includes data revealing racial or ethnic origin, political opinions, religious or philosophical beliefs, or a natural person’s sex life or sexual orientation.  For such data, consent is usually the only lawful basis for processing.   In our tests, searches for e.g. gay dating apps, muslim and christian prayer apps, diabetes, skin cancer and pregancy health apps, are used by Apple servers to select and display adverts within the search results.   Of course the user entered the search text and would like to see relevant search results, but they do not want that data used for other purposes such as advertising which they have not consented to.

\item \textit{No Consent}.
Consent is neither sought nor given for (i) the storage of non-essential cookies, adverts and other data by Apple servers, (ii) the use of sensitive personal data to target adverts, (iii) the unnecessarily intrusive tracking of how people use their iPhone.  There is no opt out from any of this data storage and processing.

\end{enumerate}

The work reported here is inevitably limited in nature.  Nevertheless, our measurements are already enough to raise concerns that Apple may be violating EU data privacy regulations.



\subsection{Mitigations}
Apple provides users with no control over the storage of cookies, adverts etc by the Apple App Store, over the use of search data to target adverts or over the tracking by Apple of how they use their iPhone.   We found that some of the user interaction data transmitted by the Apple App Store can be blocked by using a third-party firewall such as \vv{pi-hole}\footnote{https://pi-hole.net/} running on an WiFi access point to block connections to Apple servers \nolinkurl{xp.apple.com} and \nolinkurl{ca.iadsdk.apple.com}\footnote{Note that simply changing /etc/hosts on the DNS server does not work as iOS makes a DNS query for an \nolinkurl{xp.apple.com} HTTPS record and this bypasses /etc/hosts.}.   However, this requires the ability to install software on the WiFi access point, which is often not possible, and does not stop transmission using cellular connections.

\subsection{Recommendations To Apple}
We recommend that: (i) Apple provide public information on the cookies and other data that their servers store on iPhones, and obtain iPhone users consent before storing these, (ii) stop using search queries to target adverts without consent, (iii) allow users to opt out from the display of adverts within the Apple App Store, and from telemetry tracking user interactions with the App Store, and (iv) in any case such telemetry should be anonymous.   

Apple currently requires a person to login to an Apple user account before they can install an app using the Apple App Store.  This user account is directly linked to the real identity of the user, including their name, email and e.g. credit card details.  While that makes sense for paid apps, it is obviously not necessary for free apps such as the ones used in our tests.    We recommend that Apple allow users to install free apps anonymously, without requiring being logged in to an Apple account.   Since the apps that people use can be used to profile them (gay dating apps, muslim prayer apps, pregnancy tracking apps and so on), this would provide a valuable boost to the privacy of iPhone users.

\section{EU Legal Context}\label{sec:legal_short}

We report on a technical study here, not a legal one, and we are not legally qualified.  Nevertheless, the behaviour that we observe raises obvious questions regarding the EU e-Privacy Directive and perhaps also the GDPR data protection regulations (the measurements were all carried out within Ireland using handsets purchased in Ireland and so it is Irish/European data regulations that apply).   We therefore summarise the legal context below.

\subsection{Storage Of Data On A Handset}
Article 5(3) of the e-Privacy Directive is sometimes referred to as the ``cookie law''.   It was introduced in 2002\footnote{\nolinkurl{https://eur-lex.europa.eu/LexUriServ/LexUriServ.do?uri=CELEX:32002L0058:en:HTML}}, amended in 2009\footnote{\nolinkurl{https://eur-lex.europa.eu/legal-content/EN/TXT/?uri=celex:32009L0136}}  and implemented in Irish law by Statutory Instrument No. 336 of 2011\footnote{\nolinkurl{https://www.dataprotection.ie/en/who-we-are/data-protection-legislation}, accessed Jan 7th 2025.}.   

In summary, storage of data on a handset requires that: (i) information on the data storage be clear and comprehensive, prominently displayed and easily accessible, and state the purposes, and (ii) that  the user has given their consent to that use.   An exception us allowed when the storage is ``\emph{strictly necessary in order to provide an information society service explicitly requested by the subscriber or user}''\footnote{There is also a second exception ``\emph{`for the sole purpose of carrying out the transmission of a communication over an electronic communications network}'', but that is not relevant here.}.   

European guidelines further clarify that:
\begin{enumerate}
\item This regulation encompasses \emph{any} information storage on a handset, not just cookies, trackers or personally identifying/sensitive data, and that distribution of tracking links to a handset constitutes storage~\cite{edpbguidelines, edpbinterplay}.    
\item For the ``strictly necessary'' exemption to apply: (i) the user must have taken a positive action to request a service, and (ii) the service does not work if the storage is disabled~\cite{edpbcookies2012}.
\item Consent under the e-Privacy Directive has the same meaning as in the GDPR~\cite{edpbconsent},\cite[Article 2(3)]{irishlaw}, and so must be ``freely given, specific, informed and unambiguous''~\cite{edpbconsent}. 
\item Legitimate interest is not a valid legal basis for storing cookies and other data~\cite{edpbcookiebanners2023}.
\end{enumerate}

For consent to be ``freely given, specific, informed and unambiguous''~\footnote{Recital 32 Conditions for consent, \nolinkurl{https://gdpr.eu/recital-32-conditions-for-consent/}}:
\begin{enumerate}
\item Users must have a genuine choice without facing detriment for refusing consent.
\item Users must receive clear, transparent information about what data is collected and how it will be used.
 \item This information must be prominently displayed and easily accessible~\cite[Article 5(3)]{irishlaw} and should be as user-friendly as possible~\cite[Article 5(4)]{irishlaw}. 
 \item Consent must be granular, allowing users to agree to specific types of data processing.  Consent may not be ``bundled'' for multiple purposes.  
 \item Consent must be expressed through a clear affirmative action, such as clicking an ``Accept'' button.  Silence, pre-ticked boxes or inactivity do not, for example, constitute consent.
 \item  A user must be able to withdraw consent as easily as they gave it.
\end{enumerate}

\subsection{Processing of Personal Data}

Introduced in May 2018, the GDPR~\cite{gdpr} applies to the processing of personal data.   Personal data is any information that relates to an identified or identifiable living individual. Different pieces of information, which together can lead to the identification of a particular person, are considered personal data, as well as data that has been de-identified, encrypted or pseudonymised but can be used to re-identify a person~\cite{gdpr_eu}.  This includes, for example, IP addresses, cookie identifiers and advertising identifiers.   Data processing includes, amongst other things, data transmission and data storage. 

If a data controller (the person or organisation that decides why and how personal data is processed) wants to collect personal data for whatever
purpose, they need to rely on one of the six lawful grounds listed in Article 6(1) of the GDPR. Only two usually apply in the
context of mobile apps, namely: (i) consent for a defined purpose and (ii) legitimate interests.   Article 9 of the GDPR places additional obligations on ``special categories'' of data, including data revealing racial or ethnic origin, political opinions, religious or philosophical beliefs, a natural person’s sex life or sexual orientation.  For such data, consent is usually the only relevant basis for processing\footnote{The other basis are processing necessary: (i) for the purposes of carrying out obligations in the field of employment, social security and social protection law, (ii) to protect the vital interests of the data subject where they are physically or legally incapable of giving consent, (iii) in the course of legitimate activities by a foundation, association or any other not-for-profit body with a political, philosophical, religious or trade union aim, (iv) personal data manifestly made public by the data subject, (v) for the establishment, exercise or defence of legal claims, (vi) for reasons of substantial public interest, (vii) for the purposes of preventive or occupational medicine, (viii) for reasons of public interest in the area of public health, (ix) for archiving purposes in the public interest, scientific or historical research purposes.}.

Consent must be ``freely given specific, informed and unambiguous''~\cite{gdpr_eu}.  See above for a summary of what this means in practice.

Legitimate interest is an alternative to consent but requires a careful balancing exercise
between the controller’s interests in processing the personal
data and the data subjects’ interests and fundamental rights
~\cite[Article 6(1)f]{gdpr}.   The data controller must~\cite{gdpr_eu, edpblegitimateinterests,dpcguidance}:
\begin{enumerate}
\item Inform individuals about the processing when collecting their personal data.
\item Ensure the use of personal data is necessary i.e. the legitimate interests pursued cannot reasonably be achieved by other means.
\item Check that by pursuing its legitimate interests the rights and freedoms of those individuals are not seriously impacted.  This entails a balancing of the opposing rights and interests at issue which depends on the specific circumstances of the relevant processing. 
\end{enumerate}

When both the ePrivacy Directive and the GDPR apply in
a given situation, the rules of the ePrivacy Directive override the
GDPR~\cite{edpbinterplay}.   When the ePrivacy Directive requires consent
for storing data, then while consent is one option under the GDPR it becomes the only viable
one when combining both regulatory frameworks.  When the ePrivacy Directive does not require consent, e.g. when the conditions for the ``strictly necessary'' exemption are met, then when the data stored is personal data the GDPR applies and there needs to be a lawful ground, e.g. consent or valid legitimate interest, for the data storage~\cite{edpbinterplay}.

\section{Experimental Setup}
\subsection{Hardware and Software Used}
Apple iPhone 8 running iOS 16.7.11\footnote{We used a pre-A12 iPhone as an unpatchable bootloader exploit allows jailbreaking for later versions of iOS than on more recent handsets.   iOS 16.7 is, however, the latest version available since Apple have now stopped updates for pre-A12 handsets.  See \nolinkurl{https://theapplewiki.com/wiki/Jailbreak}.}, jailbroken using Palera1n v2.2.1\footnote{\nolinkurl{https://github.com/palera1n/palera1n}} and with the Ellekit\footnote{\nolinkurl{https://github.com/tealbathingsuit/ellekit}} Objective-C function hooker.  We intercepted network connections using Mitmdump 13.0.0.dev.  We bypass SSL certificate pinning within handset processes using a modified version of SSL KillSwitch\footnote{Available at \nolinkurl{https://github.com/doug-leith/poor-privacy-practices-of-the-apple-app-store/tree/main/ssl-kill-switch3-DL}}.

\subsection{Decrypting HTTPS Connections}\label{sec:mitm}
We route handset traffic via a WiFi access point (AP) that we control, configure this AP to use \vv{mitmdump} as a proxy \cite{mitmproxy} and adjust the firewall settings to redirect all WiFi HTTP/HTTPS traffic to \vv{mitmdump}  so that the proxying is transparent to the handset.   When a process running on the handset starts a new network connection, the \vv{mitmdump}  proxy pretends to be the destination server and presents a fake certificate for the target server.   This allows \vv{mitmdump} to decrypt the traffic.  It then creates an onward connection to the actual target server and acts as an intermediary, relaying requests and their replies between the app and the target server while logging the traffic. 

System processes typically carry out checks on the authenticity of server certificates received when starting a new connection and abort the connection when these checks fail.  The happens even when the mitmproxy CA cert is installed on the handset as a trusted certificate. To bypass these checks, the iPhone is jailbroken, Ellekit is installed and a custom Objective C injection script is used to bypass SSL certificate pinning within handset processes.  This script is invoked on launch of all processes that make use of the com.apple.AuthKit, com.apple.UIKit and com.apple.aps.framework frameworks and also the following processes: \nolinkurl{com.apple.softwareupdated}, \nolinkurl{com.apple.AssetCacheLocatorService}, \nolinkurl{com.apple.imfoundation.IMRemoteURLConnectionAgent}, \nolinkurl{com.apple.mobileactivationd}, \nolinkurl{com.apple.itunescloudd}, \nolinkurl{com.apple.identityservicesd},  \nolinkurl{com.apple.akd}, \nolinkurl{com.apple.itunesstored}.

\subsection{Test Protocol}

We carry out experiments with a privacy-conscious but busy/non-technical user, who when asked does not select options that share data with Apple but otherwise leaves handset settings at their default value.   We use a two-step process to jailbreak the phone and allow decryption of HTTPS traffic. Firstly, we jailbreak the phone using Palera1in with the Sileo package manager, (ii) use Sileo to download and install Ellekit and openssh, (iii) connect to the handset via ssh and install the Objective C injection script used to bypass SSL certificate pinning.  The handset has no SIM inserted.  Then we factory reset the phone and the user goes through the onboarding process without using a network connection, selecting the settings shown in Table \ref{tab:settings}.  We jailbreak the phone again using Palera1in.  Ellekit, openssh and the SSL bypass script are all left installed by the factory reset and become activated again after this second jailbreak without the need for a network connection.  The user connects to the test WiFi network and installs the mitmproxy CA cert as a trusted certificate by downloading it from the special URL http://mitm.it supported by \vv{mitmdump}, enabling it in the phone settings and setting it to be trusted.  At this point system HTTPS network connections can now be intercepted and successfully decrypted, and we record all handset network traffic.

\begin{table}[htb]
\centering
\footnotesize
\begin{tabular}{|p{3cm} | p{5cm} |}
\hline
\textbf{Onboarding screen} & \textbf{Option selected} \\
\hline
Language & "English", "Ireland"\\
Quick Start & "Set Up Manually"\\
Written and Spoken Languages & "Continue"\\
Choose a Wi-Fi Network & "Continue without wifi", "Continue"\tablefootnote{An "Unable to activate" screen is displayed.  We click "Other options", "Use Mac or PC", then connect iphone to a Macbook using a USB cable.  On the Macbook we open Apple Music, click the iPhone icon, then "Open in Finder", a popup says "Activating" and, after a delay, the iPhone moves on to next screen.  Disconnect USB cable.}\\
Data \& Privacy &"Continue"\\
Touch ID & "Set up Touch IS later"\\
Create a Passcode & "Passcode options", "Don't use passcode"\\
Apps \& Data & "Don't Transfer Apps \& Data" \\
Apple ID &"Forgot password or don't have an Apple ID", "Set up later in Settings"\\
Terms and Conditions & "Agree"\\
Update your iPhone Automatically & "Only Download Automatically"\\
"iMessage & "FaceTime", choose "Not now"\\
Location Services &  "Disable Location Services" \\
Siri & "Set Up Later in Settings"\\
Screen Time & "Set Up Later in Settings"\\
iPhone Analytics &"Don't Share" \\
Appearance & "Continue"\\
Display Zoom" &"Continue"\\
Welcome to iPhone" &"Get Started"\\
\hline
\end{tabular}
\caption{Settings selected during handset onboarding following a factory reset.}\label{tab:settings}
\end{table}

\begin{figure}[htb]
\centering
\begin{subfigure}[b]{0.22\columnwidth}
\includegraphics[width=\columnwidth]{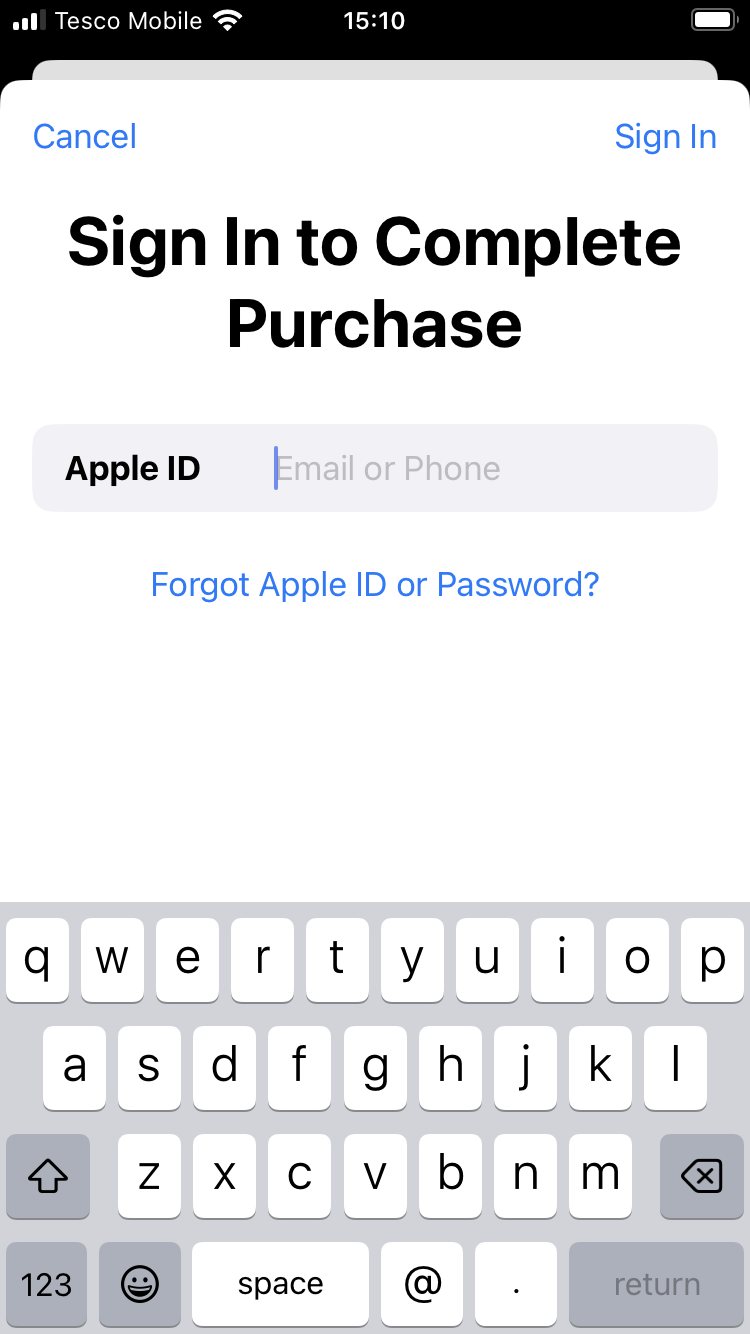}
 \caption{}
\end{subfigure}
\begin{subfigure}[b]{0.22\columnwidth}
\includegraphics[width=\columnwidth]{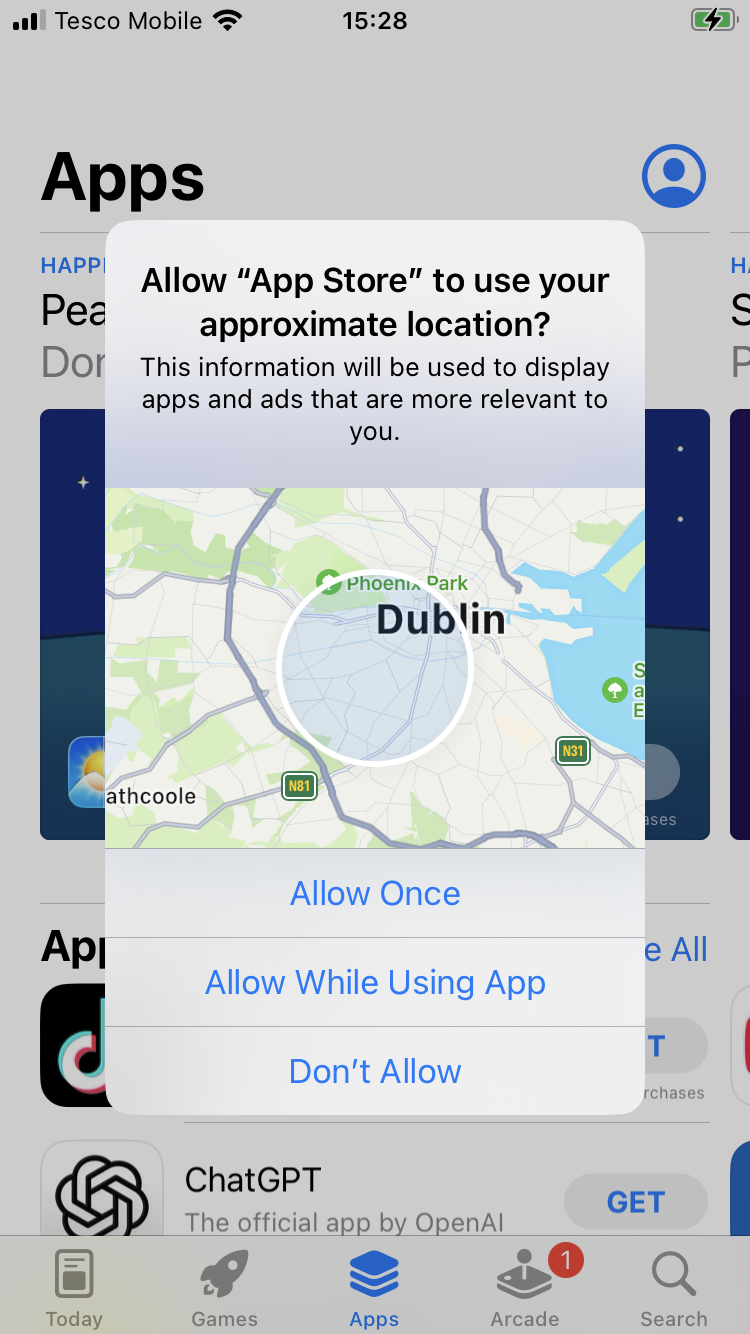}
 \caption{}
\end{subfigure}
\begin{subfigure}[b]{0.22\columnwidth}
\includegraphics[width=\columnwidth]{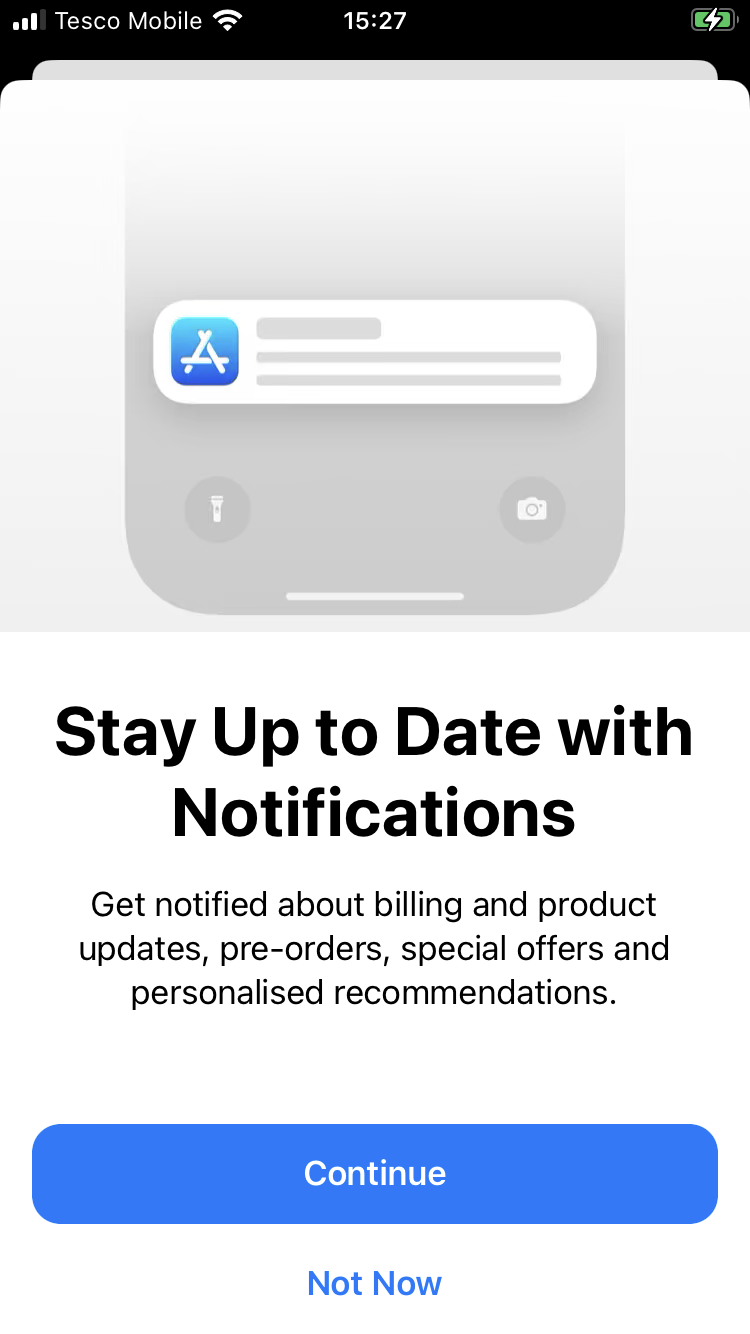}
 \caption{}
\end{subfigure}
\caption{Popups displayed when using Apple App Store. }\label{fig:popup_screenshots}
\end{figure}

\begin{figure}[htb]
\centering
\begin{subfigure}[b]{0.22\columnwidth}
\includegraphics[width=\columnwidth]{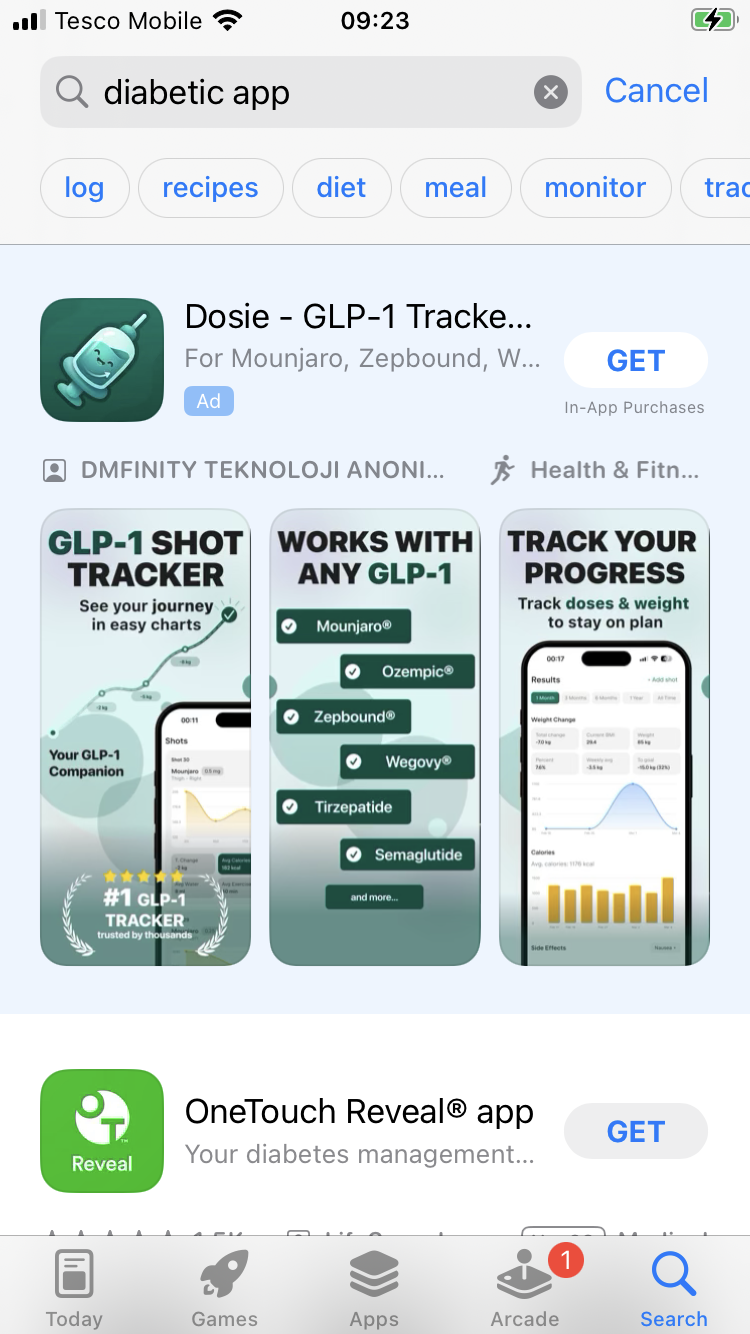}
 \caption{}
\end{subfigure}
\begin{subfigure}[b]{0.22\columnwidth}
\includegraphics[width=\columnwidth]{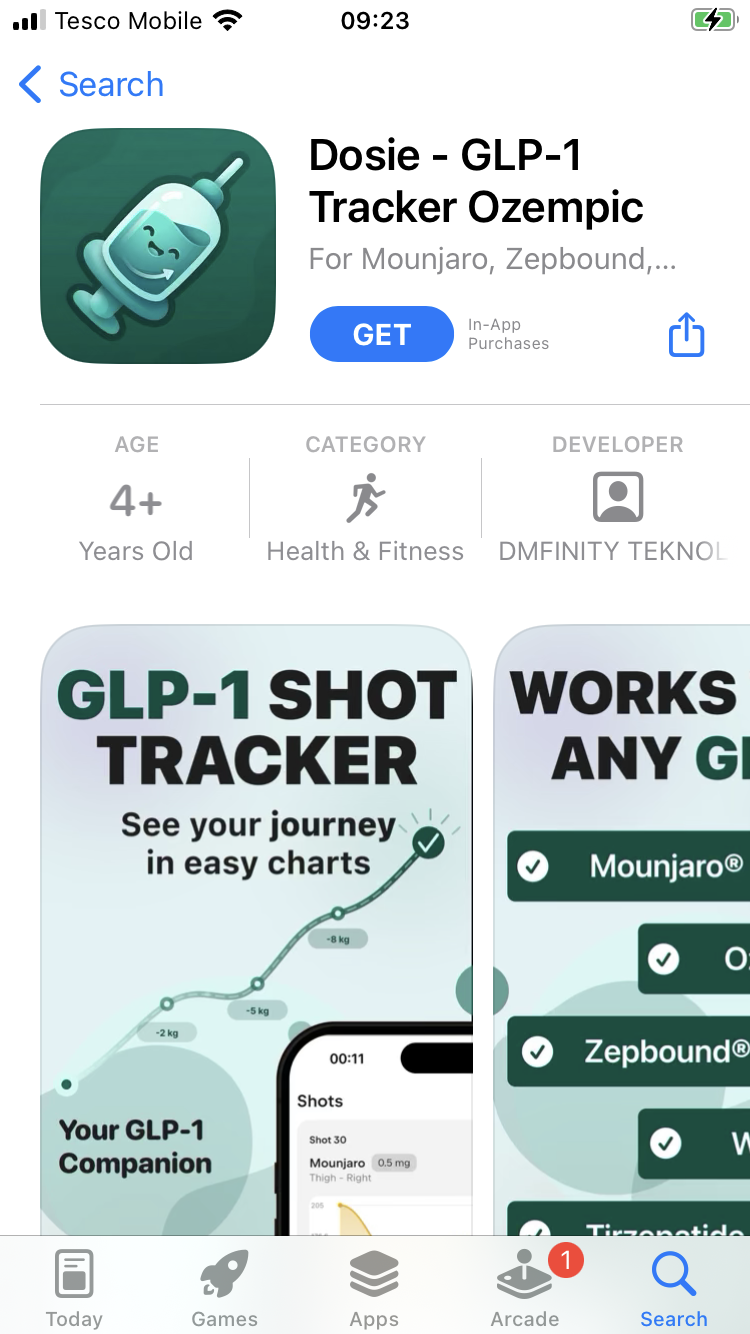}
 \caption{}
\end{subfigure}
\begin{subfigure}[b]{0.22\columnwidth}
\includegraphics[width=\columnwidth]{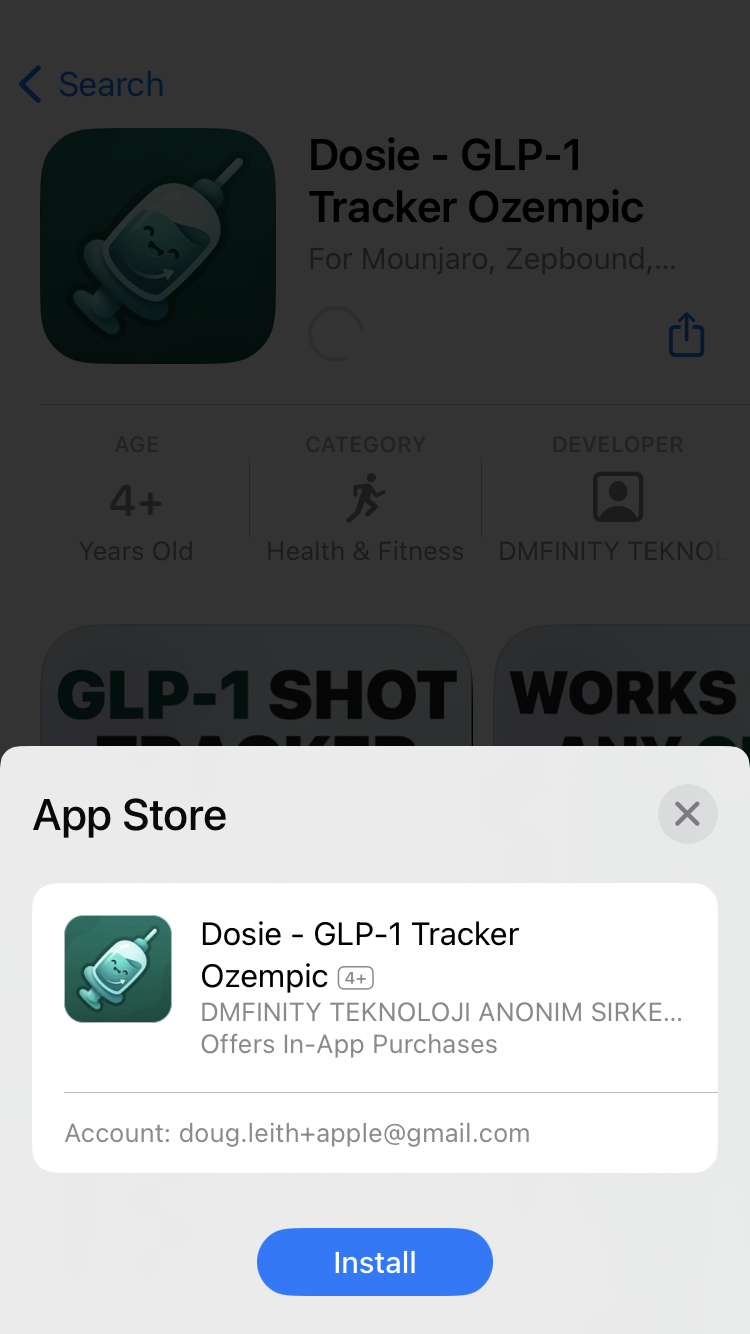}
 \caption{}
\end{subfigure}
\begin{subfigure}[b]{0.22\columnwidth}
\includegraphics[width=\columnwidth]{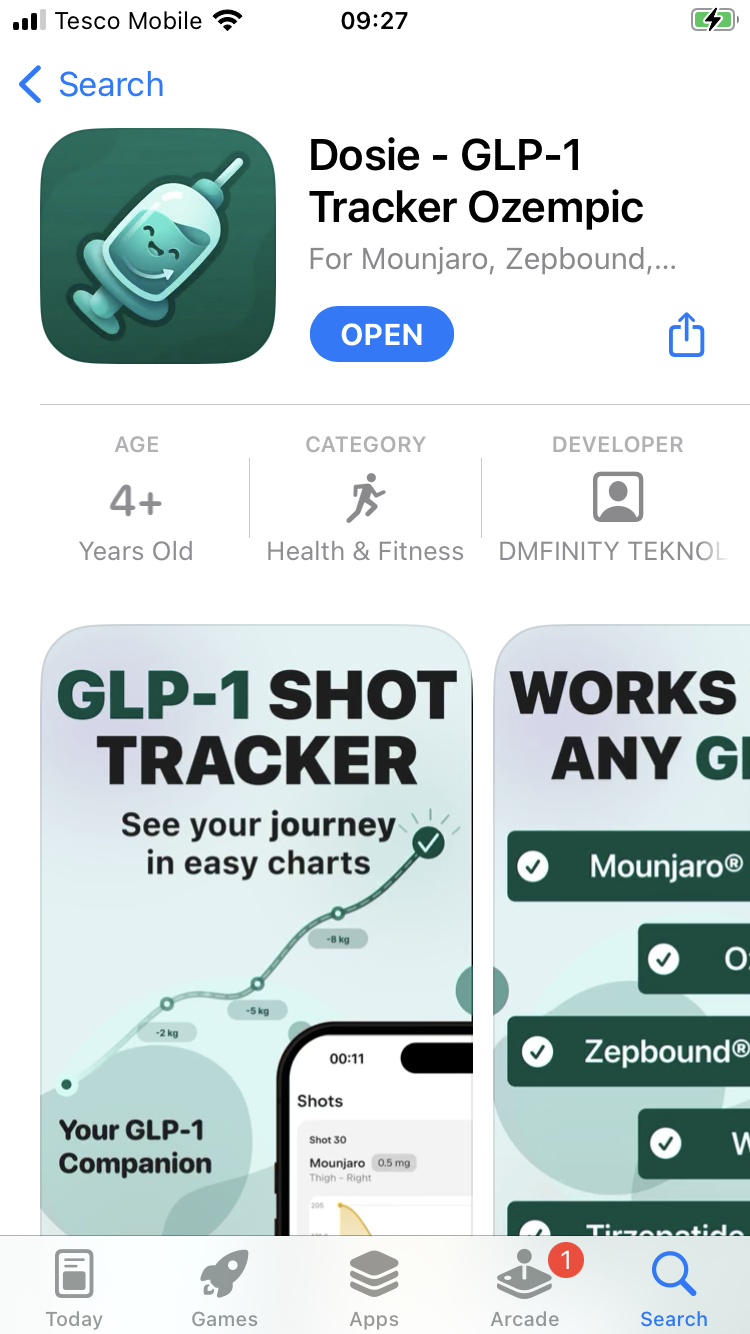}
 \caption{}
\end{subfigure}
\caption{Selected Apple App Store screenshots when searching for ``diabetes app'' and clicking on the first search result for the Dosie app (an advert). }\label{fig:search_diabetes_screenshots}
\end{figure}

The rest of the experiment is divided into two parts.  In the first part the user is not logged in to an Apple account and the handset location setting is off.  In the second part the handset location setting is turned on and the user logs in to an Apple account.  In both parts of the experiment the user opens the Apple App Store app.  
This makes a request for service by that app, but no more than that.  The Apple App Store app is of particular interest since it is the primary means by which users install other apps on the handset, and so fundamental to the usefulness of the handset.   In more detail, we carry out the following steps:

\begin{enumerate}
\item \emph{Experiment Part 1} The handset is left idle and connected to power for about 24 hours to allow time for software updating (the icons for many of the pre-installed apps are initially greyed out following the factory reset, awaiting an online check/update).  At this point, the icons for the pre-installed apps are now coloured, indictating their installation is complete, location is off and the user has never logged in to an Apple account. The test user opens the Apple App Store, does a search for ``gay dating apps'', clicks on the Hinge dating app in the results (this is the first result returned and is marked as an advert) to open the app information page, clicks ``Get'' to try to install the app, is shown the message ``sign in to complete purchase' and a request for an Apple ID', see Figure \ref{fig:popup_screenshots}(a), the user clicks ``Cancel''.   The user is shown a ``Personalised Ads screen'' popup with the buttons ``Turn on personalised ads'' and ``Turn off personalised ads'', they click ``Turn off personalised ads'', and see search results page again.   The user then clicks the ``Home''' button to close the Apple App Store.

\item \emph{Experiment Part 2}  The test user opens the Settings app, clicks ``Privacy \& Security'', ``Location Services''' and turns ``Location Services'' on.  They then open the Apple App Store.  A popup asks ``Allow "app store" to use your approximate location?'', see Figure \ref{fig:popup_screenshots}(b), the user clicks ``Don't Allow''.  The App Store then asks ``Stay Up to Date with Notifications'', see Figure \ref{fig:popup_screenshots}(c), the user clicks ``Not Now''.  As before, search for ``gay dating apps'', click on Hinge app, click ``Get'' then ``Cancel'' and return to home screen.  The test user opens the Apple App Store, clicks on the person icon at the top right, enters an Apple ID and password. A SIM is inserted into the handset, but without an active data plan. In the Apple App Store, the test user searches for ``skin cancer app'', clicks on Scanoma - Mole Check app (the first search result, an advert), clicks ``Get'', ``Install'', enters password when asked and waits for app to download and install.  Then clicks the ``Open'' button in the Apple App Store to open the newly installed app.  The test user also carries out other searches including ``muslim pro'', ``christian prayers'', ``muslim dating apps'', ``pregnancy app'', ``diabetes app'' within the  Apple App Store, clicking on the displayed results.   Figure \ref{fig:search_diabetes_screenshots}
 shows, for example, screenshots taken when searching for ``diabetes app'' and clicking on the first search result for the Dosie app (an advert).
 \end{enumerate}

Following the main experiments, we carried out an additional experiment where cookies and telemetry connections are blocked using a custom \vv{mitmdump} addon:
\begin{enumerate}
\item Cookies are deleted from all handset transmissions, except for transmissions to \nolinkurl{p52-buy.itunes.apple.com} where cookie \vv{mt-tkn-19129383683} is allowed ro be transmitted but all other cookies are deleted.
\item Transmissions to \nolinkurl{xp.apple.com} and \nolinkurl{ca.iadsdk.apple.com} are blocked, returning an HTTP 200 success code to the client but transmitting nothing to Apple servers\footnote{Note that we investigated blocking these connections by modifying the \vv{/etc/hosts} on the WiFi access point that acted as DNS server for the phone, so that these two domains resolve to 0.0.0.0.  We also manually configured the phone settings to use the WiFi access point as DNS server (the automatic settings did that anyway, but we wanted to be explicit about the user's choice).  However, we observed that the handset nevertheless continued to resolve \nolinkurl{xp.apple.com} to an Apple IP address.   It therefore appears that iOS silently overides the user's preference for DNS server when resolving some Apple domains.   This has implications for users employing popular DNS-based firewalls such as pi-hole.}.
 \end{enumerate}
The test user then opened the Apple App Store and confirmed that they could still view, search, install and open apps (i.e. the primary functions of any app store, including the Apple App Store\footnote{The Apple App Store web page \nolinkurl{https://www.apple.com/app-store/} is consistent with this, highlighting ``Get the details on every app'', ``a rich search experience'' and  ``Purchase safely and securely'' as key functions of the Apple App Store.}).   We also repeated this test using \vv{pi-hole}\footnote{https://pi-hole.net/} rather than \vv{mitmdump} to block connections to  \nolinkurl{xp.apple.com} and \nolinkurl{ca.iadsdk.apple.com} and again confirmed that the user could still view, search, install and open apps.

\subsection{Measurement Data \& Additional Material}

The full measurement data and the software tools we developed to analyse it are available on github\footnote{\nolinkurl{https://github.com/doug-leith/poor-privacy-practices-of-the-apple-app-store}.\label{foot:github}}.

\section{Cookie \& Other Data Storage By Apple}

%

\subsection{Cookies}
\subsubsection{Cookie Storage}
Table \ref{tab:cookies} summarises the cookies observed being stored on the handset, together with the connections setting the cookies, the likely purpose of each cookie and whether it is set before the user logs in to their Apple account.  Table \ref{tab:cookies_files} shows where the cookies are stored on the handset (some of the cookie values, e.g. that of the \vv{X-Dsid}, are also stored in other locations on the handset).  

\begin{table*}[htb]
\centering
\footnotesize
\begin{tabular}{| p{4cm} | p{2cm} | p{6.25cm}| p{2cm} | p{0.5cm}| p{0.5cm}| }
\hline
\textbf{Cookie} & \textbf{HTTP Domain} & \textbf{Connection That Sets Cookie} &\textbf{Likely Purpose} & \multicolumn{2}{c|}{\textbf{Local Storage}}\\
\cline{5-6}
& & & & \textbf{Expt Part 1} &\textbf{Expt Part 2}  \\
\hline
X-Dsid, pldfltcid, tv-pldfltcid & .apple.com & \nolinkurl{auth.itunes.apple.com/auth/v1/native} & user identifier & -& $\checkmark$\\
\hline
xt-b-ts-19129383683& .apple.com & \nolinkurl{auth.itunes.apple.com/auth/v1/native}, \nolinkurl{p52-buy.itunes.apple.com/WebObjects/MZBuy.woa/wa/buyProduct}& user  identifier & & $\checkmark$\\
\hline
mt-asn-19129383683, mt-tkn-19129383683 & .itunes.apple.com & \nolinkurl{auth.itunes.apple.com/auth/v1/native} & user  identifier &-& $\checkmark$\\
\hline
mz\_at\_ssl-19129383683, mz\_at0-19129383683 & .apple.com & \nolinkurl{auth.itunes.apple.com/auth/v1/native} &user  identifier &- & $\checkmark$\\
\hline 
xp\_aci & .xp.apple.com & \nolinkurl{xp.apple.com/register}, \nolinkurl{xp.apple.com/report/2/xp_amp_appstore} & analytics, user  identifier & $\checkmark$&$\checkmark$ \\
\hline
amp & {.music.apple.com}, .itunes.apple.com & \nolinkurl{auth.itunes.apple.com/auth/v1/native}, \nolinkurl{musicstatus.itunes.apple.com/v1/status} & account settings & $\checkmark$&$\checkmark$ \\
\hline
fsnw & .itunes.apple.com & \nolinkurl{entitlements.itunes.apple.com/inApps/v1/subsEntitlements?segment=News} &account settings &- & $\checkmark$\\
\hline
fstv & .itunes.apple.com &\nolinkurl{entitlements.itunes.apple.com/inApps/v1/subsEntitlements?segment=TV} &account settings& -& $\checkmark$\\
\hline
fsic & .itunes.apple.com & \nolinkurl{entitlements.itunes.apple.com/inApps/v1/subsEntitlements?segment=iCloud} & account settings& -&$\checkmark$ \\
\hline
fspc & .itunes.apple.com & \nolinkurl{entitlements.itunes.apple.com/inApps/v1/subsEntitlements?segment=Podcasts} & account settings& -& $\checkmark$\\
\hline
fsac & .itunes.apple.com & \nolinkurl{entitlements.itunes.apple.com/inApps/v1/subsEntitlements?segment=Activity} & account settings& -& $\checkmark$\\
\hline
fsas & .apple.com & \nolinkurl{entitlements.itunes.apple.com/inApps/v1/subsEntitlements?segment=AppStore} & account settings& -& $\checkmark$\\
\hline
hsaccnt & .apple.com & \nolinkurl{auth.itunes.apple.com/auth/v1/native} & account settings &- & $\checkmark$\\
\hline 
xt-src & .apple.com & \nolinkurl{auth.itunes.apple.com/auth/v1/native}, \nolinkurl{p52-buy.itunes.apple.com/WebObjects/MZBuy.woa/wa/buyProduct}& account settings & & $\checkmark$\\
\hline
itspod & .apple.com & \nolinkurl{auth.itunes.apple.com/auth/v1/native}, \nolinkurl{inappcheck.itunes.apple.com/WebObjects/MZFastFinance.woa}, \nolinkurl{p52-buy.itunes.apple.com/WebObjects/MZBuy.woa/wa/buyProduct} & account settings &- & $\checkmark$\\
\hline
xp\_ab & .apple.com & \nolinkurl{xp.apple.com/register}, \nolinkurl{xp.apple.com/report/2/xp_amp_appstore} & ? & -&$\checkmark$\\
\hline
aa & gsa.apple.com & \nolinkurl{gsa.apple.com/auth/verify/trusteddevice} & ? & -& $\checkmark$\\
\hline
vrep & .apple.com & \nolinkurl{p52-buy.itunes.apple.com/MZBuy.woa/wa/buyProduct }&? & -& $\checkmark$\\
\hline
wosid-lite& .apple.com & \nolinkurl{auth.itunes.apple.com/auth/v1/native}, \nolinkurl{p52-buy.itunes.apple.com/WebObjects/MZBuy.woa/wa/buyProduct}& ? & & $\checkmark$\\
\hline
\end{tabular}
\caption{Summary of cookies observed being stored on the handset by Apple servers.}\label{tab:cookies}
\end{table*}

\begin{table}[htb]
\centering
\begin{tabular}{|p{0.95\columnwidth}|}
\hline
\textbf{Local Storage Locations}\\
\hline
 \nolinkurl{/private/var/mobile/Library/AppleMediaServices/Accounts/0C474A6C-EF30-49A8-BD25-59372E340B28/cookies.sqlitedb}\\
 \hline
 \nolinkurl{/private/var/mobile/Containers/Data/Application/9F7F94AD-99ED-4B41-A1E0-3053A1183FDA/Library/Caches/com.apple.AppStore/Cache.db}\\ 
 \hline
\nolinkurl{/private/var/mobile/Library/Caches/com.apple.ap.adprivacyd/Cache.db}\\
\hline
\end{tabular}
\caption{Locations where Apple cookies are stored on the handset.}\label{tab:cookies_files}
\end{table}

\subsubsection{Cookie Analysis}

We could not find any public Apple documentation on these cookies or their purpose, but we can nevertheless attempt to infer whether they can act as user identifiers, hold account settings etc.   

The \vv{X-Dsid} cookie value is almost certainly the Apple Directory Services Identifier, which uniquely identifies the Apple user account\footnote{\nolinkurl{https://theapplewiki.com/wiki/DSID}}.   This value is stored in multiple places on the handset linked to the Apple user account e.g. in database \nolinkurl{/private/var/mobile/Library/Caches/com.apple.appstored/storeUser.db} where it is referred to as the \vv{account_id} (the same database entry includes the \vv{apple_id} doug.leith+apple@gmail.com).  In the response from \nolinkurl{auth.itunes.apple.com/auth/v1/native} it is referred to as the \vv{dsPersonId}.  During testing, we took measurements on an iPhone 12 as well as the iPhone 8 used in the main experiments.  On both handsets when the user is logged into the same Apple account the \vv{X-Dsid} is the same, and when logged into different accounts the \vv{X-Dsid} value is different.


The \vv{pldfltcid} and \vv{tv-pldfltcid} cookies share the same value.  We could not find reference to them within the Apple system software on the handset, so they are likely used only by the Apple servers that set and receive them rather than by services running on the handset.   During testing, we took measurements on an iPhone 12 as well as the iPhone 8 used in the main experiments.   It is likely that these act as user identifiers since these cookies are only stored after the user logs in to an Apple account, their values are the same when the user is logged into the same Apple account on different handsets and change when the user logs into a different account.

The \vv{xt-b-ts-19129383683} , \vv{mt-asn-19129383683}, \vv{mt-tkn-19129383683}, \vv{mz\_at\_ssl-19129383683}, \vv{mz\_at0-19129383683} cookies are almost certainly user identifiers since each cookie name includes the \vv{X-Dsid} value, which acts as a user identifier.  Inspection of decompiled Apple system code indicates that the \vv{mz_at_ssl-19129383683} value is an authentication token linked to the user account and used by the iTunesCloud service. The value of the \vv{xt-b-ts-19129383683} cookie is likely a timestamp, but as already noted the cookie name contains the  \vv{X-Dsid} value which acts as a user identifier.

The \vv{xp_aci} is almost certainly used for linking together data on user interactions with the Apple App Store sent to Apple server \nolinkurl{xp.apple.com} i.e. it is an analytics cookie.  Its value is changed when the user logs in or out of their Apple user account.  When the user is logged in, this cookie is sent alongside user identifiers and so can be trivially linked to these and thereby become a user identifier itself.

Cookies \vv{fsnw}, \vv{fstv}, \vv{fsic}, \vv{fspc}, \vv{faas} likely encode user account settings.  Following user login to an Apple account, \vv{fsnw}, \vv{fstv}, \vv{fsic}, \vv{fspc}, \vv{faas} are set by server \nolinkurl{entitlements.itunes.apple.com/inApps/v1/subsEntitlements} with query parameters related to Apple services (News, TV, iCloud, Podcasts, AppStore, Activity).  Cookie \vv{amp} may encode itunes settings since: (i) it is set by by server \nolinkurl{musicstatus.itunes.apple.com/v1/status}, (ii) which is referred to as \vv{getSubscriptionStatusSrv} in settings data sent by \nolinkurl{init.itunes.apple.com/bag.xml} and \nolinkurl{bag.itunes.apple.com/bag.xml}, (iii) the iTunesCloud private framework refers to \vv{getSubscriptionStatusSrv}, where it is used within methods \vv{ICMusicSubscriptionStatusRequestOperation}, \vv{ICUserAuthenticationValidationRequest}.  It is possible these settings are encoded in a manner that is linked to the user account, and so may also act as user identifiers.

The \vv{itspod} cookie is referred to in settings sent by \nolinkurl{bag.itunes.apple.com/bag.xml} e.g. 'buyProduct':  '\nolinkurl{https://{p$itspod$-}buy.itunes.apple.com/WebObjects/MZBuy.woa/wa/buyProduct}', and we later see connections to nolinkurl{p45-buy.itunes.apple.com} when \vv{itspod}=45 and to nolinkurl{p52-buy.itunes.apple.com} when \vv{itspod}=52.  The \vv{hsaccnt} cookie always has value 1 and the \vv{xt-src} cookie always has value \vv{b}.  These cookies are unlikely to act as user or device identifiers.


The purposes of the \vv{aa}, \vv{wosid-lite}, \vv{xp_ab} and \vv{vrep} cookies are not clear to us.  The \vv{aa} cookie is first set by the response from \nolinkurl{gsa.apple.com/auth/verify/trusteddevice}, and then periodically changed by later responses from \nolinkurl{gsa.apple.com/auth/verify/trusteddevice} and so may act as a session token or device identifier.  The \vv{wosid-lite} and \vv{vrep} cookies are not set until the user logs in to their Apple account, and so likely either encode account settings or act as user identifiers.

\subsubsection{No Transparency, No Consent, Not Essential, No Opt Out}
At no point during our tests was the test user given information on the cookies stored, or asked whether they consented or not to the storage of cookies by Apple servers.   To the best of our knowledge there is no public Apple documentation describing the cookies stored or their purpose(s).

None of the cookies, apart from the \vv{mt-tkn-19129383683} cookie in transmissions to \nolinkurl{p52-buy.itunes.apple.com}, are essential for the services explicitly requested by the user in our tests.   When we blocked cookies in all handset transmissions, except for \vv{mt-tkn-19129383683} in transmissions to \nolinkurl{p52-buy.itunes.apple.com}, the test user was still able to (i) open the app store, (ii) view apps, (iii) search for apps, (iv) install apps and (v) open apps.  That is, all the functionality needed for the user to find and install apps continued to work.  The cookies may or may not be needed for other functionality, but they are not essential for the services explicitly requested by the user.

There is no way for the user to opt out of this cookie storage.

\subsection{Storage of Adverts}
\subsubsection{Adverts in Search Results} \label{sec:search_ads}
In our experiments the test user carries out various searches within the Apple App Store app.   We observed that the first search result was almost always an advert, e.g. see Figure \ref{fig:search_diabetes_screenshots}(a).   These adverts are sent to the handset by Apple server \nolinkurl{amp-api-edge.apps.apple.com}\footnote{The specific endpoints we observe are \nolinkurl{amp-api-edge.apps.apple.com/v1/engagement} and \nolinkurl{amp-api-search-edge.apps.apple.com/v1/catalog/ie/search}.}.  Figure \ref{fig:adfetch} shows an example of an advert being transmitted to the handset in response to a search for ``gay dating apps''.

\begin{figure}[htb]
\centering
\begin{lstlisting}[frame=single,language=http]
GET https://amp-api-search-edge.apps.apple.com/v1/catalog/ie/search?additionalPlatforms=<...>term=!{\color{green}gay dating apps}!&with=guidedSearch%2CspellCheck
!{\color{blue}Request Headers:}!
authorization : Bearer !{\color{red}eyJ...G0Q}!
user-agent : AppStore/3.0 iOS/16.7.11 model/iPhone10,1 hwp/t8015 build/20H360 (6; dt:157) AMS/1
!{\color{blue}Response Data:}!
{'results': {'ads-result': {'data': [{'
id': '!{\color{green}595287172}!', 'type': 'apps', 'href': '/v1/catalog/ie/apps/595287172?l=en-GB' <...> 
'{"impressionId":"!{\color{red}0974ea93-b2d9-3244-ae36-369834f8a906}!",
"metadata":"!{\color{red}CiQ...2Mg==}!"<...>"installAttribution":{"adNetworkId":"com.apple.ads","campaignId":10,
"uuid":"!{\color{red}e926b22c-cba4-425a-94df-8f320d971bb9}!",
"adamId":!{\color{green}595287172}!,<...>'name': '!{\color{green}Hinge Dating App: Match \& Date}!', <...>
'extraAdInfo': '{"candidateId":"!{\color{red}a309fdf3-8ed3-4612-893f-0b6086d3ca73}!"}',
 'experimentData': {'default': {'experimentId': '!{\color{red}e972bbe3-2f13-44cc-b174-19c78e05cfb3}!', 'experimentLocale': 'en-GB'}}, 
'cppData': {'ppid': '!{\color{red}51d444a6-4ac9-491f-84f9-533a41cc328c}!
<!{\color{red}advert URLs, text, pricing}!>
'search': 
<search results>
\end{lstlisting}
\caption{Example showing transmission of an advert for the Hinge dating app to the handset by Apple server \nolinkurl{amp-api-edge.apps.apple.com} in response to a search for ``gay dating apps''.}\label{fig:adfetch}
\end{figure}


The \vv{authorization : Bearer} header value is a JWT token
that is sent to the handset by server \nolinkurl{sf-api-token-service.itunes.apple.com/apiToken}
, see Section \ref{sec:jwt} below for more details.
%
The  \vv{adamId} (also referred to as the \vv{id}) value is an app identfier\footnote{\nolinkurl{https://developer.apple.com/documentation/apple_ads/appasset}}, in the case of Figure \ref{fig:adfetch} this identifies the Hinge dating app.   The \vv{impressionId} likely uniquely identifies the advert view (or ``impression'').   The \vv{metadata} value is a base64 encoded protobuf whose first field is the \vv{impressionId} value and whose other fields contain binary values (perhaps crytographic signatures or authorization tokens).  It is not clear what the  \vv{uuid} value represents.   The \vv{experimentId} value is likely used for A/B testing of adverts.  It is sent alongside user interaction telemetry to Apple server \nolinkurl{xp.apple.com} (see below), allowing user behaviour in response to adverts to be tracked and linked to different advert settings/choices.  Table \ref{tab:advert_files} shows some of the locations where these downloaded advert identfiers are stored on the handset.

In Figure \ref{fig:adfetch}, the search results which follow the adverts contain the \vv{adamId}, which is an app identifier, but none of the other identifiers sent with the adverts.  We see similar behaviour in other Apple App Store connections downloading app data, i.e. the \vv{impressionId}, \vv{metadata} and \vv{experimentId} are only downloaded and stored for apps which are adverts.

\begin{table}[htb]
\centering
\footnotesize
\begin{tabular}{| p{3cm} | p{5cm} |}
\hline
\textbf{Advert Content} & \textbf{Local Storage Location}\\
\hline
\vv{adamId}, \vv{id} & \nolinkurl{Engagement/journeys/app.db}\tablefootnote{The full path is \nolinkurl{/private/var/mobile/Library/Caches/com.apple.AppleMediaServices/Engagement/journeys/app.db}}, \nolinkurl{Engagement/analytics/app.db}, \nolinkurl{Metrics/AppStore/metrics.sqlitedb}, \nolinkurl{com.apple.AppStore/fsCachedData/CB8E8F74-226F-41BD-9221-3BF0690E80B3}\tablefootnote{The full path is \nolinkurl{/private/var/mobile/Containers/Data/Application/9F7F94AD-99ED-4B41-A1E0-3053A1183FDA/Library/Caches/com.apple.AppStore/fsCachedData/CB8E8F74-226F-41BD-9221-3BF0690E80B3}}, \nolinkurl{com.apple.AppStore/Cache.db}\tablefootnote{The full path is \nolinkurl{/private/var/mobile/Containers/Data/Application/9F7F94AD-99ED-4B41-A1E0-3053A1183FDA/Library/Caches/com.apple.AppStore/Cache.db}}\\
\hline
\vv{impressionId} & \nolinkurl{Engagement/journeys/app.db}, \nolinkurl{Engagement/analytics/app.db}, \nolinkurl{Metrics/AppStore/metrics.sqlitedb}\\
\hline
\vv{metadata} & \nolinkurl{Engagement/journeys/app.db}, \nolinkurl{Engagement/analytics/app.db}, \nolinkurl{Metrics/AppStore/metrics.sqlitedb}\\
\hline
\vv{ppid}, \vv{pageCustomId} & \nolinkurl{Engagement/journeys/app.db}, \nolinkurl{com.apple.AppStore/fsCachedData/CB8E8F74-226F-41BD-9221-3BF0690E80B3}, \nolinkurl{com.apple.AppStore/Cache.db}\\
\hline
\vv{experimentId} & \nolinkurl{Engagement/journeys/app.db}, \nolinkurl{com.apple.AppStore/fsCachedData/CB8E8F74-226F-41BD-9221-3BF0690E80B3}\\
\hline
 Advert URLs, text, pricing & \nolinkurl{com.apple.AppStore/fsCachedData/CB8E8F74-226F-41BD-9221-3BF0690E80B3}\\
\hline
\end{tabular}
\caption{Locations where Apple advert data downloaded in Figure \ref{fig:adfetch} is stored on the handset.}\label{tab:advert_files}
\end{table}

In Figure \ref{fig:adfetch} the user is not logged in to their Apple account.  When they are logged in, we observe that in addition the \vv{DSID} and \vv{x-apple-adsid} user identifiers are sent as HTTP request headers, also the \vv{x-apple-i-md-m} device identifier and multiple cookies (including \vv{mz_at0-19129383683}, \vv{wosid-lite}, \vv{xt-b-ts-19129383683}, \vv{pldfltcid}, \vv{tv-pldfltcid}, \vv{xp_ab}, \vv{fasas}, \vv{mz_at_ssl--19129383683}, \vv{X-Dsid}, \vv{vrep}).

\subsubsection{Adverts Pro-actively Downloaded}

\begin{figure}[htb]
\centering
\begin{subfigure}[b]{0.22\columnwidth}
\includegraphics[width=\columnwidth]{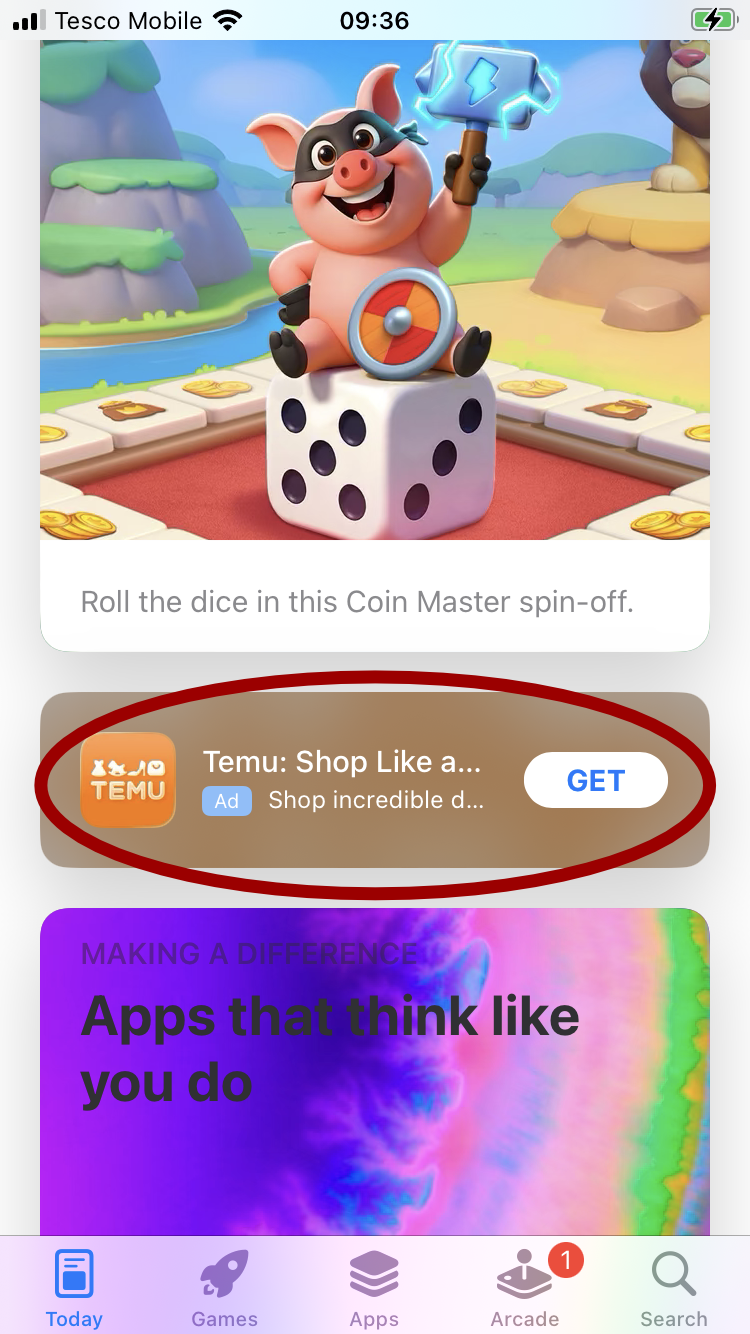}
 \caption{}
\end{subfigure}
\begin{subfigure}[b]{0.22\columnwidth}
\includegraphics[width=\columnwidth]{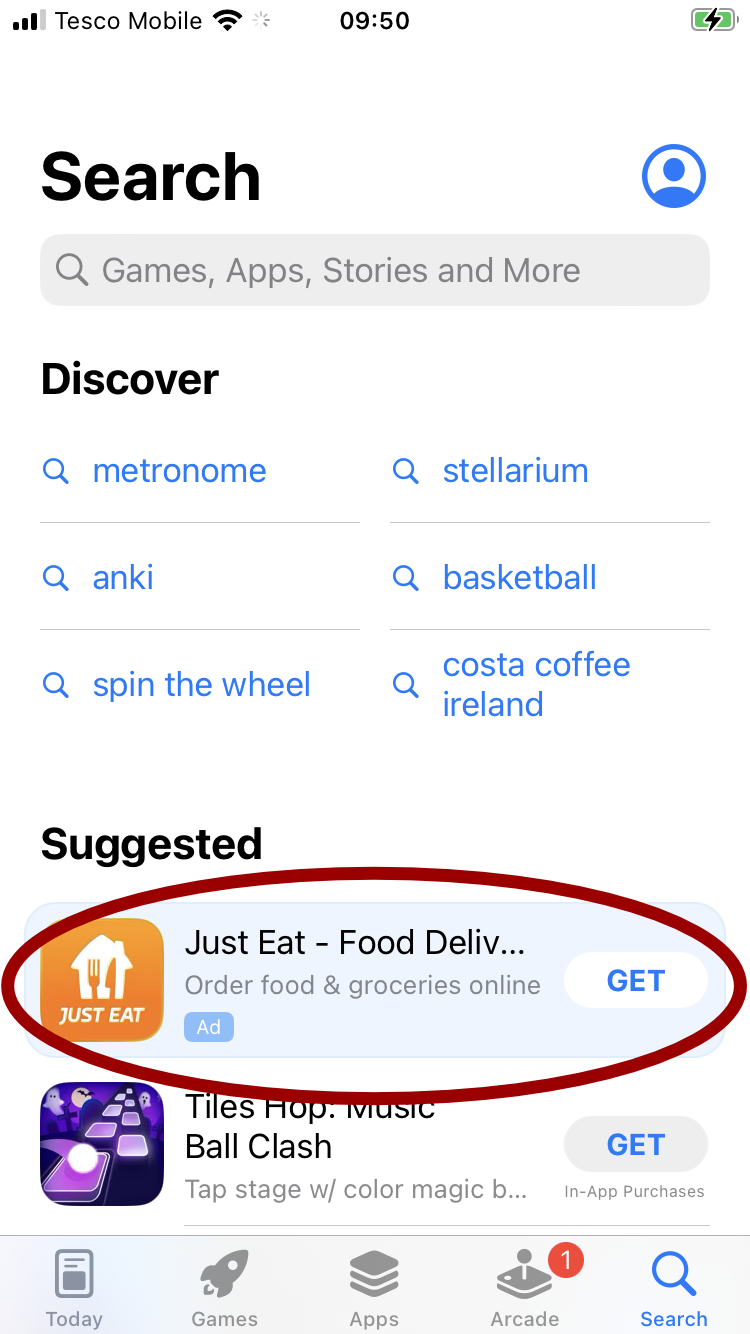}
 \caption{}
\end{subfigure}
\caption{Screenshots showing adverts displayed on the Apple App Store home page and on the search page (before any search is made). Red ovals are added by us to highlight the adverts.}\label{fig:proactive_ads}
\end{figure}

Adverts are also pro-actively downloaded and stored by the Apple App Store.  For example, Figure \ref{fig:proactive_ads} shows adverts displayed on the Apple App Store home page (after the user has opened the App Store but not done anything else) and on the search page (after the user has clicked on the search icon at the bottom of the page, but before any search has been made).  Figure \ref{fig:adfetch2} shows an advert for the DropChef food delivery app\footnote{The Apple App Store page for the DropChef can be viewed at \nolinkurl{https://apps.apple.com/ie/app/dropchef/id1574947021}. Observe that the id value at the end of the URL matches the adamId value in the response shown in Figure \ref{fig:adfetch2}.} being sent to the handset in response to a request for adverts sent to Apple servers by the Apple App Store. 

\begin{figure}[htb]
\centering
\begin{lstlisting}[frame=single,language=http]
POST https://amp-api-edge.apps.apple.com/v1/engagement/ie/ads?placement=search-landing&platform=iphone&sparseLimit=0
!{\color{blue}Request Headers:}!
user-agent : AppStore/3.0 iOS/16.7.11 model/iPhone10,1 hwp/t8015 build/20H360 (6; dt:157) AMS/1
authorization : Bearer !{\color{red}eyJr...3MQ}!
!{\color{blue}POST Body:}!
"data": {"adRequest": <...>"clientId":" !{\color{red}AC515964-3260-447D-AA21-C86A3876135F}!"<...>
!{\color{blue}Response Data (decoded):}!
{'results': {'ads': [
{'id': '!{\color{green}1574947021}!', 'type': 'apps', 'href': '/v1/catalog/ie/apps/!{\color{green}1574947021}!', 
'meta': {'adResult': '{
		"impressionId":"!{\color{red}6b74929a-c2e5-4681-8180-dc23bab8191f}!",
		"metadata":"!{\color{red}CiQ...kgB}!",
		"privacy":"EkB...2Fk",
		"journeyRelay":{"adGroupId":1677628104,"campaignId":1677547825},
		"clientRequestId":"3BDF3C04-733D-4279-95E2-A562001748B2",
		"installAttribution":{
			"adNetworkId":"com.apple.ads","campaignId":20,
			"uuid":"!{\color{red}a76eba1c-0a12-47dd-a803-caecbbf4e2a4}!",
			"adamId":!{\color{green}1574947021}!,
			"timestamp":1771423683730, "signature":"MEU...03d", "version":"3.0",
			"sourceAppAdamID":!{\color{green}915061776}!},
		"ttl":32400,"rank":1,"AdType":"Regular", "triggers":[{"experimentId":"bs","treatmentId":"t1"},{"experimentId":"e668","treatmentId":"t0|r4"}]
<more adverts>
\end{lstlisting}
\caption{Example showing an advert being pro-actively download by the Apple App Store. }\label{fig:adfetch2}
\end{figure}

The \vv{adamId},  \vv{metadata} (and so \vv{impressionId}), \vv{adGroupId}, \vv{campaignId}, \vv{sourceAppAdamID}, \vv{clientRequestId}, \vv{type}, \vv{href} and \vv{timestamp} values are stored on the handset in file \vv{69A30B-BCE9-4D31-A424-3F7B5FBB4333.d} within folder \vv{/private/var/mobile/Library/Application Support/com.apple.ap.promotedcontentd/SFS/y/D7}.   

%

\subsubsection{No Transparency, No Consent, Not Essential, No Opt Out}
At no point during our tests was the test user given information on the adverts stored, or asked whether they consented or not to the storage of adverts by Apple servers.  

Storage of adverts is self-evidently not essential for the services explicitly requested by the user in our tests.  In the Apple App Store the test user would still be able to view apps, search for apps, install apps and open apps without adverts being stored on their handset by the Apple App Store.

There is no way for the user to opt out of this advert storage.

\subsection{Storage of Identifiers \& Tokens}\label{sec:identifiers}

\subsubsection{Storage of User Identifiers}

When the test user logs in to their Apple account, Figure \ref{fig:login} shows that several identifiers are transmitted to the handset by Apple server \nolinkurl{gsa.apple.com} (this connection is made at the end of the login process, immediately after the user enters a two-factor authentication verification code).

\begin{figure}[htb]
\centering
\begin{lstlisting}[frame=single,language=http]
POST https://gsa.apple.com/auth/verify/trusteddevice/securitycode
!{\color{blue}Request Headers:}!
Cookie: site=IRL; aa=66B0E97FEB760851B895CCEC31493C53
X-Mme-Device-Id : !{\color{red}cceed3efd045a9b4cee8dfcf7b05a852164e888c}!
X-Apple-I-MD-M : !{\color{red}iiYeXFEeMERWMF+8YQOFHPLOIyMyOEWhQNdtJdEeaIvZrYsyLoukwpag+u6NHE+CHTE0DkfZMDo7P42L}!
X-Apple-Identity-Token : MDA...VPQ==
User-Agent : AppStore/1 CFNetwork/1410.1 Darwin/22.6.0
!{\color{blue}POST Body:}!
<...><key>securityCode.code</key><string>379525</string><key>serverInfo</key><dict><key>phoneNumber.id</key><string>1</string><...>
!{\color{blue}Response Headers:}!
X-Apple-Identity-Token : Y29t...DQ5
X-Apple-GS-Token : Y29..NDk=
X-Apple-HB-Token : Y29...0OQ==
X-Apple-DSID : !{\color{red}19129383683}!
X-Apple-I-CK : ODhiOWQ4NzU5OTExZTI1ZjBkMjk2ZWQ4YzQyZjViZA==
X-Apple-I-Account-Name : !{\color{red}doug.leith+apple@gmail.com}!
X-Apple-Session-Key : mKv/sQX8BHMQIVrFmOy51PTLyqyGJ+PZsLd5UnU2iXs=
X-Apple-Alternate-Id : !{\color{red}000191-10-d0cd9af9-d7b6-4b63-ac3b-71dd4e81c73f}!
X-Apple-Encrypted-Session-Key : MTE...RnBI
X-Apple-PE-Token : Y29...wMA==
!{\color{blue}Response Data:}!
<xmlui action="dismiss" />
\end{lstlisting}
\caption{Apple server \nolinkurl{gsa.apple.com} sending identifiers to the handset after the user logs in to their Apple account.}\label{fig:login}
\end{figure}

\begin{table}
\centering
\footnotesize
\begin{tabular}{| p{1.5cm} | p{1.4cm} | p{4.75cm}|}
\hline
\textbf{Identifier} & \textbf{Other Names\tablefootnote{HTTP headers, query parameters etc used in network connections to refer to the same value}} & \textbf{Local Storage Location}\\
\hline
Account Name/Email & X-Apple-I-Account-Name, x-apple-adsid & \nolinkurl{/private/var/mobile/Library/Accounts/Accounts3.sqlite}\\
\hline
Account Id/DSID & X-Apple-DSID, x-dsid, dsId, userId &  \nolinkurl{/private/var/mobile/Library/Caches/com.apple.appstored/storeUser.db}, \nolinkurl{/private/var/mobile/Library/Preferences/com.apple.itunescloud.daemon.plist}, \nolinkurl{/private/var/mobile/Library/AppleMediaServices/Accounts/0C474A6C-EF30-49A8-BD25-59372E340B28/cookies.sqlitedb}, \nolinkurl{/private/var/mobile/Library/Caches/com.apple.AppleMediaServices/Metrics/AppStore/metrics.sqlitedb}\\
\hline
Alternate DSID & X-Apple-Alternate-Id, x-apple-adsid & \nolinkurl{/private/var/mobile/Library/Caches/com.apple.appstored/storeUser.db}, \nolinkurl{/private/var/mobile/Library/Accounts/Accounts3.sqlite}\\
\hline
\end{tabular}
\caption{User identifiers downloaded from Apple servers and stored on the handset.  For identifiers such as the DSID which are stored in many locations, only a selection of these are shown. }
\label{tab:user_ids}
\end{table}

The \vv{X-Apple-I-Account-Name} value is the email associated with the Apple user account.  The \vv{X-Apple-DSID} value is likely the Apple Directory Services Identifier, which uniquely identifies the Apple user account.   The \vv{X-Apple-Alternate-Id} value is likely a second identifer that uniquely identifies the Apple user account.  Table \ref{tab:user_ids} summarises the identfiers and where they are stored on the handset.  

\begin{figure}[htb]
\centering
\begin{subfigure}[b]{0.22\columnwidth}
\includegraphics[width=\columnwidth]{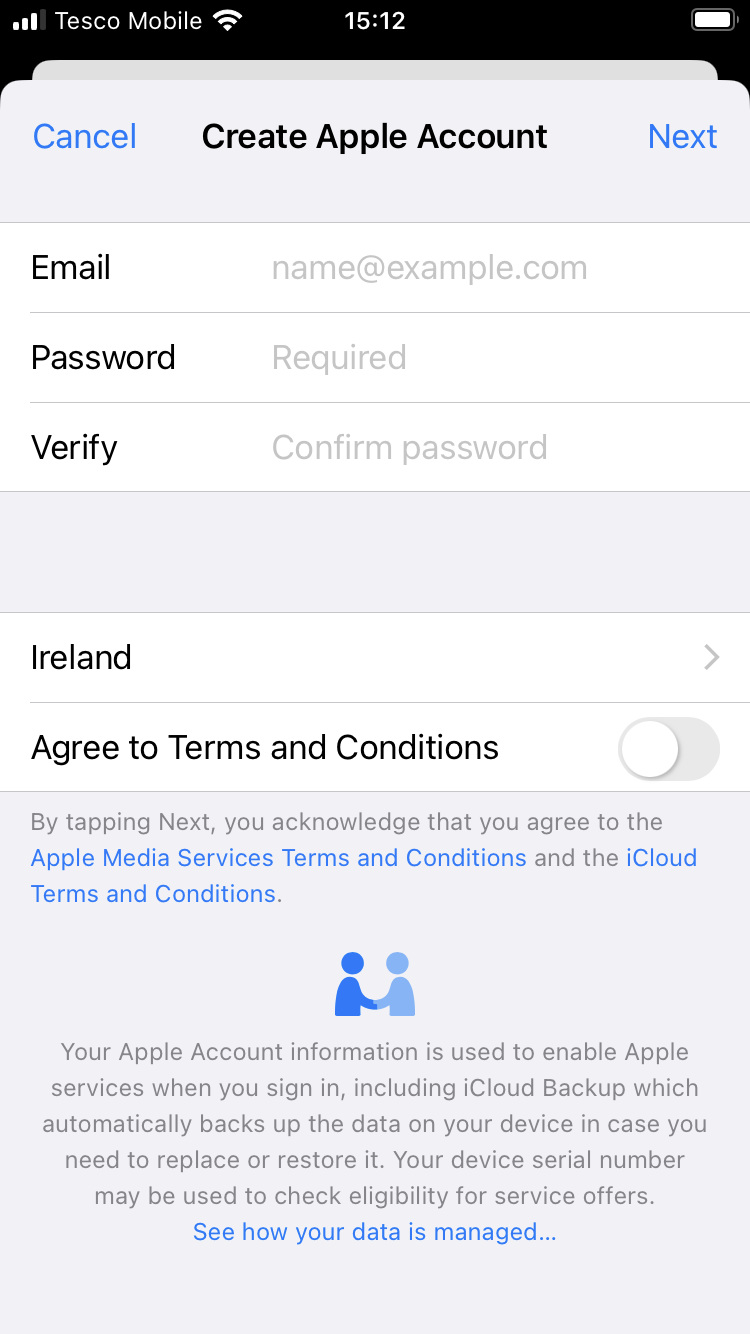}
 \caption{}
\end{subfigure}
\begin{subfigure}[b]{0.22\columnwidth}
\includegraphics[width=\columnwidth]{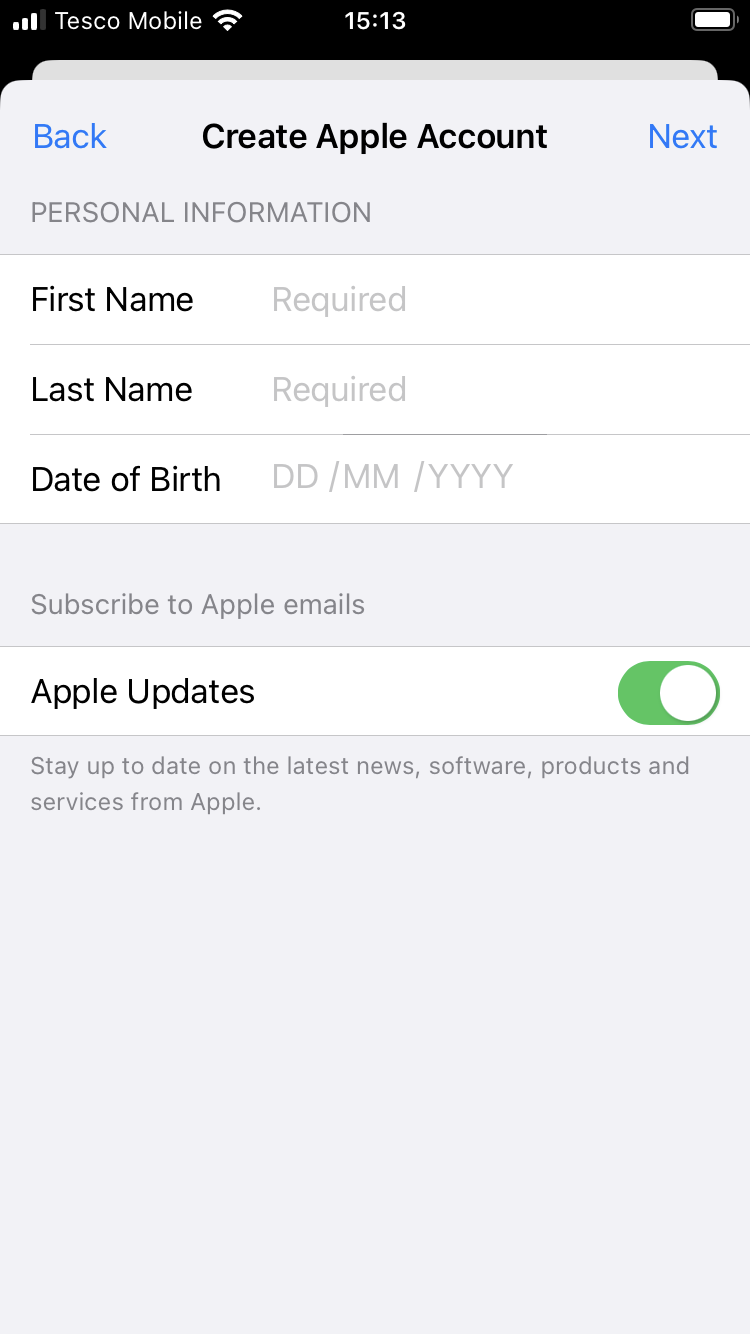}
 \caption{}
\end{subfigure}
\begin{subfigure}[b]{0.22\columnwidth}
\includegraphics[width=\columnwidth]{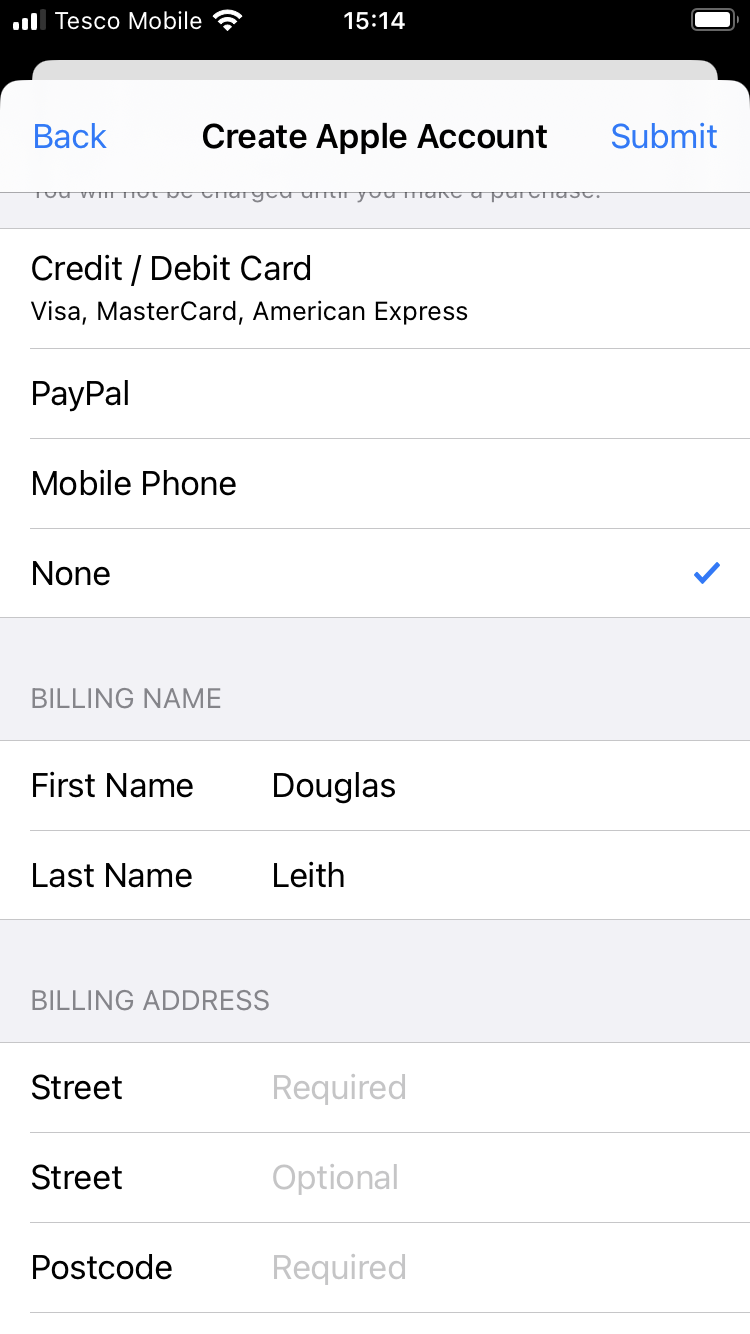}
 \caption{}
\end{subfigure}
\begin{subfigure}[b]{0.22\columnwidth}
\includegraphics[width=\columnwidth]{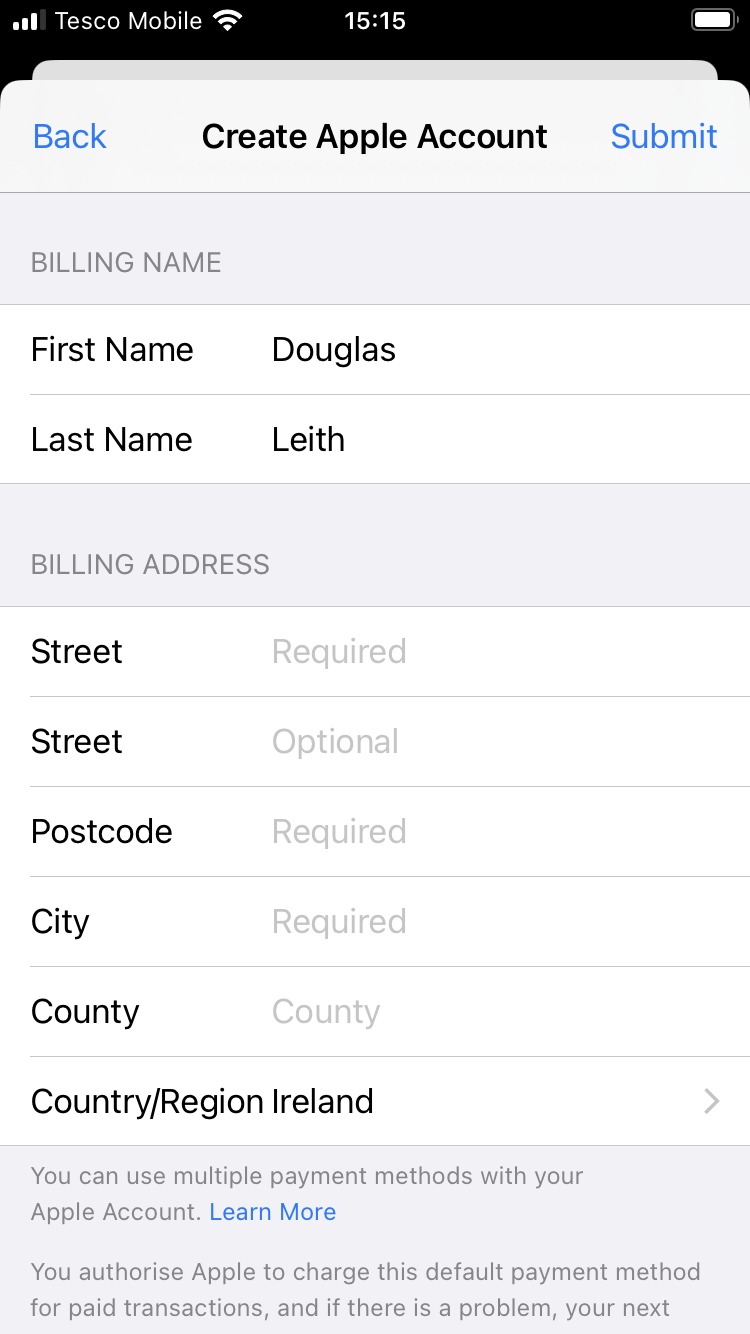}
 \caption{}
\end{subfigure}
\caption{Selected screenshots when creating a new Apple user account. }\label{fig:account_screenshots}
\end{figure}

Figure \ref{fig:account_screenshots} shows that the Apple user account is directly linked to the real identity of the user.  This figure presents screenshots taken when creating a new Apple user account.  It is necessary to enter the user's email, full name, date of birth and address (the address is required even when ``None'' is selected for billing).  It is not possible to proceed without providing these details.


\subsubsection{Storage of Authorization Tokens}\label{sec:jwt}
Several network connections transmit \vv{authorization : Bearer} header tokens, which presumably act to authenticate the connections.   These tokens are in JWT format\footnote{\nolinkurl{https://www.jwt.io/}} and are sent to the handset by the response from \nolinkurl{sf-api-token-service.itunes.apple.com/apiToken}, see for example Figure \ref{fig:jwttoken}.  

\begin{figure}[htb]
\centering
\begin{lstlisting}[frame=single,language=http]
GET https://sf-api-token-service.itunes.apple.com/apiToken?clientClass=apple&clientId=!{\color{green}appstore}!&os=iOS&osVersion=16.7.11&productVersion=3.0&version=2
!{\color{blue}Request Headers:}!
cookie : mz_at0-!{\color{red}19129383683}!=AwQ...QJuI=, wosid-lite=BdZ...TIg, pldfltcid=f5f...052, xt-b-ts-!{\color{red}19129383683}!=1772790091263, itspod=52, mz_at_ssl-!{\color{red}19129383683}!=AwU...Z70=, mt-tkn-!{\color{red}19129383683}!=Aks...cAA=, xp_abc=2b0ASLE01, X-Dsid=!{\color{red}19129383683}!, tv-pldfltcid=f5f...052, amp=+b4u...mqE=, xp_ab=1#KV...s01
x-apple-amd-m : !{\color{red}iiYe...OBV}!
x-apple-i-md-m : !{\color{red}iiY...42L}!
x-apple-adsid : !{\color{red}000191-10-d0cd9af9-d7b6-4b63-ac3b-71dd4e81c73f}!
user-agent : AppStore/3.0 iOS/16.7.11 model/iPhone10,1 hwp/t8015 build/20H360 (6; dt:157) AMS/1
!{\color{blue}Response Data:}!
{'token': '!{\color{red}eyJ...nvg}!', 'clientId': '!{\color{green}appstore}!'}
\end{lstlisting}
\caption{Example showing Apple App Store downloading a JWT token from Apple server \nolinkurl{sf-api-token-service.itunes.apple.com} when the user is logged in to their Apple account.}\label{fig:jwttoken}
\end{figure}

\begin{table*}[tb]
\centering
\footnotesize
\begin{tabular}{| p{4cm} | p{7cm} | p{3.25cm} | p{2cm} |}
\hline
\textbf{clientId} & \textbf{Connections using token} & \textbf{User-agent} & \textbf{Content}\\
\hline
\nolinkurl{appstore} & \nolinkurl{amp-api-edge.apps.apple.com/v1/catalog/ie/apps/}, \nolinkurl{amp-api-edge.apps.apple.com/v1/recommendations/ie}, \nolinkurl{amp-api-search-edge.apps.apple.com/v1/catalog/ie/search/suggestions}, \nolinkurl{amp-api-search-edge.apps.apple.com/v1/catalog/ie/search}, \nolinkurl{amp-api-edge.apps.apple.com/v1/editorial/ie/editorial-items} & AppStore & App Information, Search results, Adverts\\
\hline
\nolinkurl{com.apple.ap.promotedcontentd} &\nolinkurl{amp-api-edge.apps.apple.com/v1/engagement/ie/ads} & AppStore & Adverts\\
\hline
\nolinkurl{com.apple.manage.subscriptions} & \nolinkurl{amp-api-edge.apps.apple.com/v1/engagement/ie/upsell/marketing-items } & AppStore & Adverts for Apple products\\
\hline
\nolinkurl{com.apple.storekit.prod} & \nolinkurl{amp-api-edge.apps.apple.com/v1/catalog/ie/in-apps} & \nolinkurl{com.apple.storekitd} & Subscription offers for installed apps\\
\hline
\nolinkurl{com.apple.amp.journeys} &\nolinkurl{amp-api.media.apple.com/v1/engagement/ie/journeys} & \nolinkurl{com.apple.amsengagementd} & Analytics, User behaviour\\
\hline
\end{tabular}
\caption{Apple services that download tokens from \nolinkurl{sf-api-token-service.itunes.apple.com}.  Also shown are the connections in which the tokens are subsequently transmitted, the user-agent header in those connections (which indicates the app making the connection) and a summary of the content.}\label{fig:jwt_conns}
\end{table*}


In Figure \ref{fig:jwttoken} the user is logged into their Apple account and the request headers contain both user identifiers (e.g. DSID, \vv{x-apple-adsid}) and device identifiers (e.g. \vv{x-apple-amd-m}, \vv{x-apple-i-md-m}).   Tokens are fetched even when the user is not logged in to an Apple account, in which case the request headers only contain a device identifier (\vv{x-apple-amd-m}).   The token in the response is stored on the handset in file \nolinkurl{Containers/Data/Application/9F7F94AD-99ED-4B41-A1E0-3053A1183FDA/Library/Caches/com.apple.AppStore/Cache.db}

In addition to tokens downloaded by the Apple App Store itself, Table \ref{fig:jwt_conns} shows other Apple services associated with the Apple App Store that also download tokens from \nolinkurl{sf-api-token-service.itunes.apple.com}.  The tokens downloaded by \vv{com.apple.ap.promotedcontentd} and \vv{com.apple.manage.subscriptions} are used in network connections solely related to advertising.  The tokens downloaded by \vv{com.apple.amp.journeys}  are solely used in network connections recording user interactions with the Apple App Store (see Section \ref{sec:tracking}).    None of these are essential for the service explicity requested by the user, namely use of the Apple App Store to search for and install apps.  The tokens downloaded by \vv{appstore} are used in network connections that do provide these requested services, but in addition are also used to deliver adverts.

\subsubsection{No Transparency, No Consent For Storage Of Tokens For Advertising And User Tracking, No Opt Out}

In our tests the user logs in to their Apple account so that they can install apps using the Apple App Store.  This is required because the Apple App Store prevents installation of apps without logging in and alternative app stores are not available.  Storage of Apple account user identifiers, such as the Account name/email and Id/DSID, on the handset is probably reasonable to provide the service explictly requested by the user, namely log in to allow installation of apps.   

However, Apple goes further than this.   The user identifiers are also used by \vv{com.apple.ap.promotedcontentd} and \vv{com.apple.manage.subscriptions} to download and store authorization tokens that are used in network connections solely related to advertising, while the tokens downloaded and stored by \vv{com.apple.amp.journeys}  are solely used in network connections recording user interactions with the Apple App Store.    None of these are essential for the service explicity requested by the user.  The tokens downloaded by \vv{appstore} are used in network connections that do provide these requested services, but in addition are also used to deliver adverts.  That is, they appear to serve two purposes, only one of which was explicitly requested by the user.   At no point during our tests was the test user given information on the tokens stored, or asked whether they consented or not to the storage of tokens used by Apple for advertising and user tracking.  

There is no way for the user to opt out of this storage of tokens for advertising and analytics.

\section{Use Of Sensitive Personal Data To Target Adverts}

\begin{figure}[htb]
\centering
\begin{subfigure}[b]{0.22\columnwidth}
\includegraphics[width=\columnwidth]{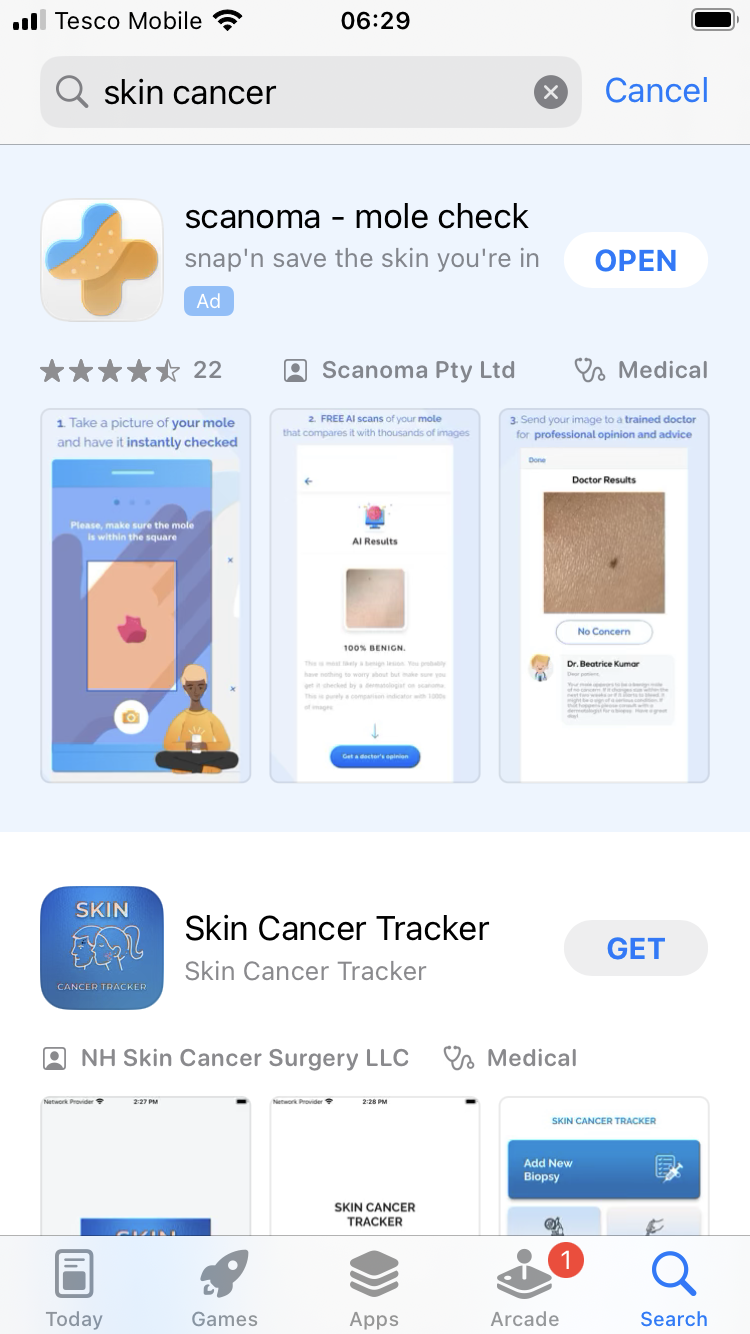}
 \caption{}
\end{subfigure}
\begin{subfigure}[b]{0.22\columnwidth}
\includegraphics[width=\columnwidth]{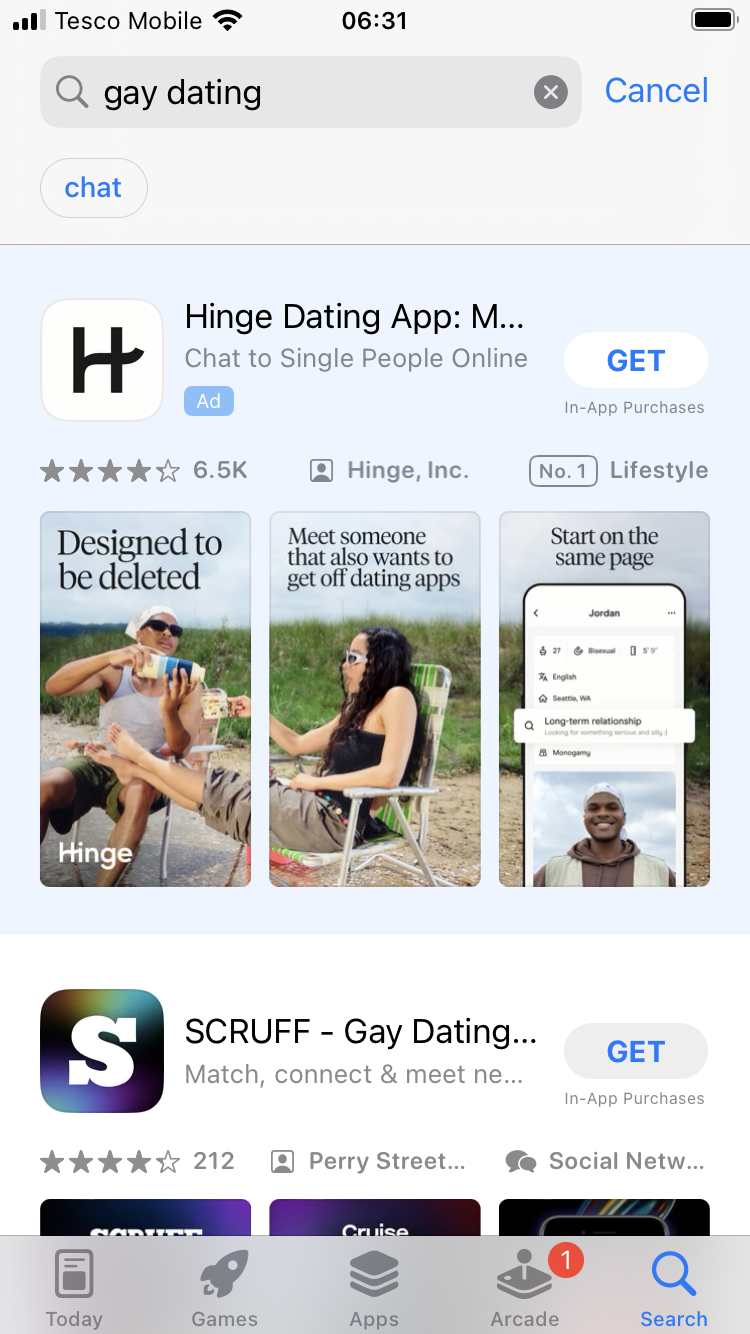}
 \caption{}
\end{subfigure}
\begin{subfigure}[b]{0.22\columnwidth}
\includegraphics[width=\columnwidth]{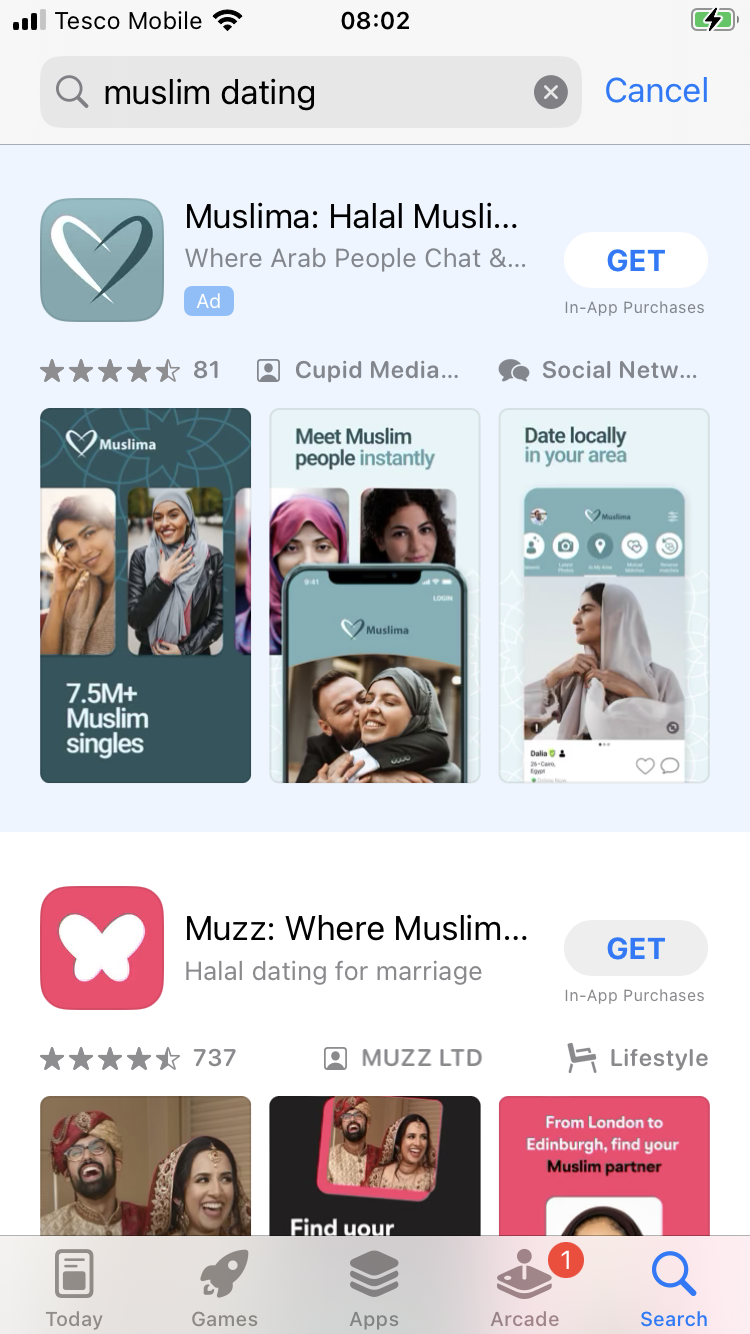}
 \caption{}
\end{subfigure}
\begin{subfigure}[b]{0.22\columnwidth}
\includegraphics[width=\columnwidth]{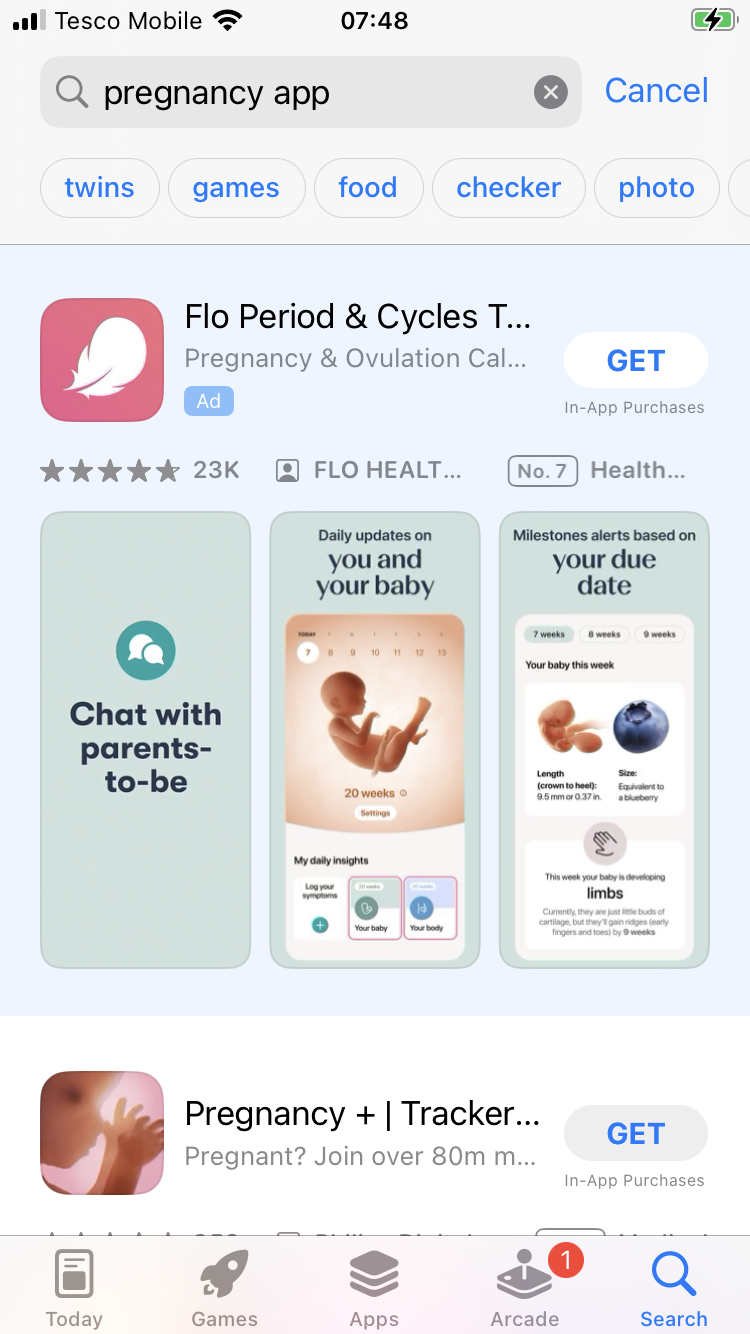}
 \caption{}
\end{subfigure}
\caption{Screenshots of search results for a variety of searches on sensitive topics.  Observe that the first search result is always an advert related to the search text entered by the user.}\label{fig:ad_screenshots}
\end{figure}

Figure \ref{fig:ad_screenshots} shows that when the test user searches for sensitive topics related to health, sexuality and religion the first search result is an advert targeted to the sensitive topic.  Figure \ref{fig:adfetch} shows that the search text is sent to Apple server \nolinkurl{amp-api-search-edge.apps.apple.com/v1/catalog/ie/search} alongside user and device identifiers. The search can therefore be linked to the user and is personal data.   In response, the Apple server sends adverts to the handset that are clearly selected based on this personal data.

\subsection{No Consent For Use Of Sensitive Personal Data For Advertising, No Opt Out}
No consent is sought or given for use of this sensitive personal data to target advertising.   
There is no way for the user to opt out of advertising, and no way for the user to opt out of this use of their data for advertising.

On the Privacy \& Security > Apple Advertising > View Ad Targeting Information screen within the Settings app on the handset, Apple states that ``relevant ads may be selected using your search query'' even when the Personalized Ads setting is off (as it is in our tests).   However, this notice buried several screens deep within the Settings apps is hardly transparent and certainly does not constitute consent under EU data privacy regulations.     In the App Store > App Store \& Arcade Privacy > App Store screen within the Settings app on the handset, Apple states that ``You can turn off personalisation features in the App Store ... by turning off Personalised Recommendations for your Apple ID.  On iOS and iPadOS devices, go to Settings > [your name] > Media \& Purchases > View Account  then tap Personalised Recommendations.''  We did this, but in App Store search results we continue to see adverts that are clearly selected based on sensitive personal data (unchanged from Figure \ref{fig:ad_screenshots}).

Note that the issue here is the use of sensitive personal data for advertising.  Of course the user entered the search text and would like to see relevant search results, but they do not want that data used for other purposes such as advertising which they have not consented to.

\section{Tracking of User Interactions}\label{sec:tracking}

\subsection{Tracking Information Sent To {xp.apple.com}}

When a user interacts with the Apple App Store app, the App Store sends details of the user interactions to Apple servers. Table \ref{fig:tracking_summary} summarises the data sent to Apple server  \nolinkurl{xp.apple.com} when the user (i) searches for ``muslim pro'' (a prayer app), (ii) clicks on the first search result (which is an advert for the ``Muslim: Ramadan 2026, Athan'' app) to view the app page, (iii) installs the app and (iv) opens it.  It can be seen that the Apple App Store transmits details of pages and adverts viewed (``impressions'''), searches typed, buttons clicked, when the user opens and closes the app etc.    Table \ref{fig:tracking_summary} also shows some of the user and device identifiers transmitted alongside this interaction data, allowing it to be linked to the specific handset and user.   See the Additional Material for more details on the data transmitted to \nolinkurl{xp.apple.com}.  



%

\begin{table*}[tb]
\centering
\footnotesize
\begin{tabular}{| p{1.75cm} | p{1.25cm} | p{3cm}|   p{1.1cm} | p{0.7cm} | p{0.9cm} | p{0.55cm} | p{1.5cm} | p{0.65cm} |  p{1.7cm} |}
\hline
\textbf{eventTime (ms)}& \textbf{eventType} & \textbf{Further details} & \multicolumn{7}{c|}{\textbf{Identifiers}}\\ \cline{4-10}

	         &                  &                         & \textbf{x-apple-i-md-m, x-apple-adsid, x-dsid headers}  & \textbf{xp\_aci cookie} & \textbf{dsId, userId, clientId} & \textbf{iAdId} & \textbf{iAd-ImpressionId, iAdMetadata} & \textbf{pageId} &\textbf{dataSetId/ pageCustomId}\\
\hline
1772720953165 & impressions & search landing page, with app adverts & $\checkmark$ & $\checkmark$ & $\checkmark$ & $\checkmark$ & $\checkmark$ & $\checkmark$ & $\checkmark$\\
1772720954289 & impressions & search landing page, with app adverts & $\checkmark$& $\checkmark$ & $\checkmark$ & $\checkmark$ & $\checkmark$ & $\checkmark$ & $\checkmark$ \\
1772720954359 & search & input "m" & $\checkmark$& $\checkmark$ & $\checkmark$  & & & &\\
1772720954565 & search & input "mu" & $\checkmark$& $\checkmark$ & $\checkmark$ & & & &\\
1772720955038 & search & input "mus" & $\checkmark$& $\checkmark$ & $\checkmark$ & & & &\\
$\cdots$ && & & & & & &&\\
1772720959568 & search &search term "muslim pro" & $\checkmark$& $\checkmark$ & $\checkmark$  & & & &\\
1772720962513 & click & press "Search" button & $\checkmark$ &$\checkmark$ &  $\checkmark$ & & & & \\
1772720985169 & page & search results page& $\checkmark$& $\checkmark$ & $\checkmark$ & $\checkmark$ & $\checkmark$ & $\checkmark$ & $\checkmark$\\
1772720985271 & media & play video  & $\checkmark$& $\checkmark$ & $\checkmark$ & $\checkmark$ & & $\checkmark$ & $\checkmark$\\
1772720990151 & impressions & advert "Muslim: Ramadan 2026, Athan"& $\checkmark$& $\checkmark$ & $\checkmark$ & $\checkmark$ & $\checkmark$ & $\checkmark$ & $\checkmark$ \\
1772720998794 & click & advert "Muslim: Ramadan 2026, Athan" & $\checkmark$& $\checkmark$ & $\checkmark$ & $\checkmark$ & $\checkmark$ & $\checkmark$ & $\checkmark$ \\
1772720998831 & media & play video & $\checkmark$ & $\checkmark$& $\checkmark$ & $\checkmark$ & & $\checkmark$ & $\checkmark$\\
1772720998934 & impressions & advert "Muslim: Ramadan 2026, Athan" & $\checkmark$& $\checkmark$ & $\checkmark$ & $\checkmark$ & $\checkmark$ & $\checkmark$ & $\checkmark$ \\
1772720998961& account & & $\checkmark$ &$\checkmark$ &  $\checkmark$  & &  & &\\
1772720999573 & media & play video, page "Muslim: Ramadan 2026, Athan" &$\checkmark$ &$\checkmark$ &  $\checkmark$ & $\checkmark$ & & $\checkmark$ & $\checkmark$ \\
1772720999969 & pageChange &  &$\checkmark$ &$\checkmark$ &  $\checkmark$ & $\checkmark$ & $\checkmark$& $\checkmark$ & \\
$\cdots$ && & & & & & &&\\
1772721034516 & click & press "Get" app button, &$\checkmark$ &$\checkmark$ &  $\checkmark$ & $\checkmark$ & $\checkmark$ & $\checkmark$ & $\checkmark$  \\
$\cdots$ && & & & & & &&\\
1772721231650 & dialog & dialogType:"SignIn", enterPasswordPressed, authenticate, result:"success"& & $\checkmark$ &  & & $\checkmark$ & &\\
$\cdots$ && & & & & & &&\\
1772721269287 & dialog & message: "Save password for Free Items?"&  &$\checkmark$ &  & & & &\\
$\cdots$ && & & & & & &&\\
1772721366483 & click & press "Open" app button & $\checkmark$ &$\checkmark$ & $\checkmark$ & $\checkmark$ & $\checkmark$ & $\checkmark$ & $\checkmark$  \\
$\cdots$ && & & & & & &&\\
1772721369893 & exit & &  $\checkmark$ &$\checkmark$ & $\checkmark$ & & & & \\
\hline
\end{tabular}
\caption{Events transmitted to \nolinkurl{xp.apple.com/report/2/xp_amp_appstore} when the user searches for ``muslim pro'' in Apple App Store, clicks on the first search result (which is an advert for the ``Muslim: Ramadan 2026, Athan'' app) to view the app page, installs the app and opens it.  The {x-dsid}, {dsId} and {x-apple-adsid} are user identifiers, the {x-apple-i-md-m} is a device identifer (see Table~\ref{tab:device_ids}).}\label{fig:tracking_summary}
\end{table*}

\subsection{Tracking Information Sent To {p52-buy.itunes.apple.com}}
When the user clicks the ``Get'' button for an app, two connections are made to Apple server \nolinkurl{p52-buy.itunes.apple.com}.  
Figure \ref{fig:p52_get} shows some of the data sent in the first connection, which includes the advertising identifiers \vv{impressionId} and \vv{adamId} (the app is an advert embedded in the search results) as well as user identifiers, including the \vv{DSID}, and the ``muslim pro'' search term used to arrive at the app page.   The second connection sends similar data, and in addition a hex string ``afds'' which encodes a ``Basic Attestation System'' certificate which contains the handset IMEI, ECID, SEID, Bluetooth and WiFi MAC addresses.

\begin{figure}[htb]
\centering
\begin{lstlisting}[frame=single,language=http]
POST https://p52-buy.itunes.apple.com/WebObjects/MZBuy.woa/wa/buyProduct?guid=!{\color{red}cceed3efd045a9b4cee8dfcf7b05a852164e888c}!
!{\color{blue}Request Headers:}!
X-Apple-I-MD-M : !{\color{red}iiYeXFEeMERWMF+8YQOFHPLOIyMyOEWhQNdtJdEeaIvZrYsyLoukwpag+u6NHE+CHTE0DkfZMDo7P42L}!
X-DSID : !{\color{red}19129383683}!
X-Apple-ADSID : !{\color{red}000191-10-d0cd9af9-d7b6-4b63-ac3b-71dd4e81c73f}!
X-Apple-iAd-Id : !{\color{red}AC515964-3260-447D-AA21-C86A3876135F}!
Cookie : X-Dsid=!{\color{red}19129383683;}! xt-src=b; xp_ab=1#K...s01; xt-b-ts-!{\color{red}19129383683}!=1771423226373; wosid-lite=6Eg...BflE=; countryVerified=1; pldfltcid=f5f...b052; fspc=AAA...hi9Y=; fsic=AAAA...JpA==; fstv=AAA...FZw==; vrep=CIP...RAA; fsnw=AAA...RVA==; fsas=AAA...hu/tA==; hsaccnt=1; mz_at_ssl-!{\color{red}19129383683}!=AwU...2V90=; utsdc=skrdc; amp=vUO...AtTo=; tv-pldfltcid=f5f5...052; fsac=AAA...iCNg==; itspod=52
User-Agent : AppStore/3.0 iOS/16.7.11 model/iPhone10,1 hwp/t8015 build/20H360 (6; dt:157) AMS/1
!{\color{blue}POST Body:}!
<...>
<key>adam-id</key><integer>!{\color{red}1137963102}!</integer>
<key>impression-id</key><string>!{\color{red}5ba839e6-9209-4dbf-9701-6ad799012dd2}!</string>
<key>mtSearchTerm</key><string>!{\color{green}muslim pro}!</string>
<...>
\end{lstlisting}
\caption{Data transmitted to \nolinkurl{p52-buy.itunes.apple.com} when the user clicks on the "Get" button in App Store for the ``Muslim: Ramadan 2026, Athan'' app.  This click is also transmitted separately to \nolinkurl{xp.apple.com}, as shown in Table \ref{fig:tracking_summary}.}\label{fig:p52_get}
\end{figure}

\subsection{Tracking Information Sent To {ca.iadsdk.apple.com}}
When the user clicks the ``Open'' button for an app, Figure \ref{fig:openclick} shows the data sent to Apple server \nolinkurl{ca.iadsdk.apple.com}.  This includes the advertising identifiers \vv{metadata} (which includes the \vv{impressionId}) and \vv{adamId}.  The ``toroId'' is referred to in connections to \nolinkurl{xp.apple.com} as the ``iAdId'' and in connections to \nolinkurl{amp-api-edge.apps.apple.com/v1/engagement/ie/ads} as the ``clientId''.  Since the same value is transmitted in all of these connections, it can easily be used to link them together.

\begin{figure}
\centering
\begin{lstlisting}[frame=single,language=http]
POST https://ca.iadsdk.apple.com/adserver/attribution/v2
!{\color{blue}Request Headers:}!
User-Agent : iPhone10,1; iOS 16.7.11-20H360; com.apple.ap.promotedcontentd; 143449-2,29
!{\color{blue}POST Body:}!
{'attributionMetadataExistsOnDevice': False, 'toroId': '!{\color{red}AC515964-3260-447D-AA21-C86A3876135F}!', 'purchaseTimestamp': '2026-03-05T14:35:17Z', 'adamId': !{\color{red}1137963102}!, 'searchAdClickTimestamp': '2026-03-05T14:29:58Z', 'adMetadata': '!{\color{red}CiR...djI=}!', 'attributionDownloadType': 0, 'developmentApp': False, 'anonymousDemandId': '8D994912-FBF0-4718-A84E-D803BC8895B2', 'bundleId': ''!{\color{green}com.muslim.assist}!', 'attributionKey': '!{\color{red}LeD...vEc=}!'}
\end{lstlisting}
\caption{Data transmitted to \nolinkurl{ca.iadsdk.apple.com} when the ``Muslim: Ramadan 2026, Athan'' app is first opened.}\label{fig:openclick}
\vspace{-0.5cm}
\end{figure}

\subsection{Tracking Information Sent To {se-edge.itunes.apple.com}}
When an app is downloaded to the handset, Figure \ref{fig:se_get} shows the data sent to Apple server \nolinkurl{se-edge.itunes.apple.com}.  The data sent includes user and device identifiers and the app \vv{adamId}.

\begin{figure}[htb]
\centering
\begin{lstlisting}[frame=single,language=http]
POST https://se-edge.itunes.apple.com/WebObjects/MZStoreElements.woa/wa/buyButtonMetaData?cc=ie&plistPage=true&type=software
!{\color{blue}Request Headers:}!
x-apple-i-md-m : !{\color{red}iiYeXFEeMERWMF+8YQOFHPLOIyMyOEWhQNdtJdEeaIvZrYsyLoukwpag+u6NHE+CHTE0DkfZMDo7P42L}!
x-apple-amd-m : !{\color{red}iiYeXFEeMERWMF+8YQOFHPLOIyMPhy9whwnCJR3oHQpPrmkshyvSitq/wv6Ao0uQ4OPjPU43AapTrOBV}!
x-dsid : !{\color{red}19129383683}!
x-apple-adsid : !{\color{red}000191-10-d0cd9af9-d7b6-4b63-ac3b-71dd4e81c73f}!
cookie : X-Dsid=!{\color{red}19129383683;}! xt-src=b; xp_ab=1#K...s01; xt-b-ts-!{\color{red}19129383683}!=1771423226373; wosid-lite=6Eg...BflE=; countryVerified=1; pldfltcid=f5f...b052; fspc=AAA...hi9Y=; fsic=AAAA...JpA==; fstv=AAA...FZw==; vrep=CIP...RAA; fsnw=AAA...RVA==; fsas=AAA...hu/tA==; hsaccnt=1; mz_at_ssl-!{\color{red}19129383683}!=AwU...2V90=; utsdc=skrdc; amp=vUO...AtTo=; tv-pldfltcid=f5f5...052; fsac=AAA...iCNg==; itspod=52
user-agent : AppStore/3.0 iOS/16.7.11 model/iPhone10,1 hwp/t8015 build/20H360 (6; dt:157) AMS/1
!{\color{blue}POST Body:}!
ids=!{\color{red}1137963102}!
\end{lstlisting}
\caption{Data transmitted to \nolinkurl{se-edge.itunes.apple.com} when the ``Muslim: Ramadan 2026, Athan'' app is downloaded.}\label{fig:se_get}
\end{figure}

\subsection{Not Anonymous, Not Essential, No Consent}
The user interaction data transmitted is not anonymous since it is tagged with user and device identifiers, and so is personal data.  The data sent is not needed to provide the service requested by the user: when transmission of data to Apple servers \nolinkurl{xp.apple.com} and \nolinkurl{ca.iadsdk.apple.com} is blocked the user can still open the Apple App Store, view, search, install and open apps (the primary functions of any app store, including the Apple App Store).   No consent is sought or given for the transmission of this user interaction data to Apple servers.  

Apple's ``App Store \& Privacy'' document\footnote{\nolinkurl{https://www.apple.com/ie/legal/privacy/data/en/app-store/}} is a complex document and we are not lawyers.  Nevertheless, we note that this document states that the legal basis for data processing are (i) ``for performance of your contract with Apple, as necessary for providing the service, and to comply with our legal obligations'', (ii) consent and (iii) Apple's legitimate interests.  Regarding legitimate interests, personal data is processed for the following purposes: (i) ``Display relevant content, assess whether your device has enough space to download requested content, and facilitate automatic downloads of updated content'', (ii) ``help improve and optimize the App Store and other Apple products and services for you and others'', (iii) ``Send you offers via email and notifications that may be of interest to you, and to create segments to serve you ads in the App Store, Apple News, and Stocks'', (iv) ``Process refund requests'', (v) ``Provide reporting to developers on purchases in their apps'', (vi) ``Prevent fraud and other malicious activity''.  This personal data includes ``Your use of the App Store app, such as your interactions with content you search for, view, download, and purchase, as well as App Store notifications and messages and in-app notifications and messages''.   

The user interaction telemetry that we see transmitted to Apple servers is not necessary for providing the service requested by the user, and consent is not sought.  Presumably, therefore, legitimate interest is the legal basis for this telemetry.  However, telemetry is not relevant to purposes (i) and (iv), while for purposes (ii) and (v) anonymous data is sufficient i.e. there is no need for the telemetry data to be tagged with user and device identifiers.  Regarding (iii), in Ireland those purposes surely require consent (e.g. see~\cite{dpc_faq}) i.e. legitimate interest is not a valid legal basis (the Apple documents states that data processing for legitimate interest only occurs ``[w]here applicable local law provides'').   That leaves (vi) which is a catch-all purpose that lacks transparency and for which processing of detailed telemetry data, including searches for sensitive topics (religious beliefs, health, sexual orientation etc), is unlikely to be essential.

\section{Limitations}
The work reported here is inevitably limited in nature.  Many different Apple connections are made, even when a handset is lying idle, and we have analysed only a subset of them.  In addition, in our experiments we only make fairly minimal use of the handset.  This means that the data collection and storage which we report here is likely an underestimate.

In our tests we use a jailbroken handset.  This is necessary to allow decryption of the HTTPS network traffic, as discussed in Section \ref{sec:mitm}.   It is possible that the behaviour of a jailbroken handset differs from that of a non-jailbroken one.  However, we have seen no evidence of this and in any case we would expect jailbreaking to \emph{decrease} app activity such as data storage, e.g. due to security concerns when jailbreakting is detected, in which case the data collection and storage that we report here would be an underestimate.  

In our tests we also assume a privacy conscious user who answers ``no'' when asked to share data with Apple and who uses no Apple services other than the Apple App Store.  Again, this means that the data collection and  storage which we report here is likely an underestimate.

Probably the main limitation of our work is the use of an older version of iOS.  This is unavoidable since there are no jailbreaks for later versions.   It may be that the behaviour of later versions of iOS differs from what we report here, but there is little reason to believe that.  For example, adverts are still displayed in Apple App Store search results on the latest version of iOS, and still selected based on sensitive search queries.

\section{Related Work}
\subsection{Tracking using browser cookies} 
A number of large-scale measurement studies have used web browsers to document the cookies stored by web sites, how they can be used to track user activity across the web and how behaviour has changed over time.  See for example~\cite{10.1145/2872427.2882991, 10.1145/2736277.2741679,10.1145/3176246,203894,OmniCrawl, 8002896, 10.1145/3501247.3531545}.   There has also been related work on the back and forth between web cookie blockers and the development of methods for circumventing these blockers e.g. use of browser fingerprinting, changing to using first-party cookies when third-party cookies are blocked~\cite{10.1145/3442381.3449837,8706882,10.1145/3386040,10.1145/3576915.3616586,7872467,217648}.

\subsection{Impact of legislation on browser cookie behaviour}  
Several web browser measurement studies investigate the impact on web cookie behaviour of the introduction of legislation, in particular the EU e-Privacy Directive, GDPR and the California Consumer Privacy Act (CCPA).   Typically these look at whether cookies are set before consent is given by the user, since this is easy to check automatically at scale, e.g. see~\cite{10.1007/978-3-030-15986-3_17,10.1145/3708515,10.1145/3442381.3450056}.   There have also been a number of studies analysing cookie consent notices displayed to users, e.g.~\cite{4yearscookies,  ndss2019} and investigations by the Irish Data Protection Commissioner~\cite{irishdpc} and European Data Protection Board~\cite{edpbcookies}.


\subsection{Automated web cookie policy enforcement} 
More recently, in response to the tiresome nature of consent notices, several browser-based tools have been proposed for automated cookie policy enforcement and evaluated using measurement studies e.g. ~\cite{277190} develops a browser extension, called CookieBlock, that uses machine learning to enforce GDPR cookie consent at the client while~\cite{10.1007/978-3-031-28486-1_26} evaluates the BannerClick with Priv-Accept tools.



\subsection{Network traffic analysis}
A number of large-scale measurement studies have used network traffic analysis to detect transmissions to tracking services by mobile apps, and to investigate the data transmitted.  These use rooted phones and a man-in-the-middle attack to intercept and decrypt HTTPS traffic leaving the phone.  Typically, apps are installed from an app store, run for a short period with no user interaction, and the network traffic recorded and analysed for transmission of sensitive information such as location and personal identifiers.  There is no analysis of cookies or data stored by the apps.  See, for example,~\cite{274431,274614,iphones,infocomgaen21,securecom21,snooping21,securecom22,narseo18}.

\subsection{Mobile OS-level telemetry}
While there has been a fair number of measurement studies on app telemetry, mostly telemetry sent by third-party SDKs, there has been much less work on data transmitted by the mobile OS itself.  The earliest work appears to be~\cite{infocomgaen21} with follow-on studies by~\cite{securecom21,snooping21,securecom22}.  The only previous work on Apple's iOS is~\cite{securecom21} but there is no analysis of cookies or data stored.   Recently,~\cite{LEITH2026104850} has looked at the cookies and other data stored by Google apps, and in particular Google Play Services, on Android.

\bibliographystyle{IEEEtran}
\bibliography{bibfile}

\section*{Additional Material}
\setcounter{section}{0}
\setcounter{subsection}{0}
\setcounter{subsubsection}{0}

\begin{table*}[htb]
\centering
\begin{tabular}{| p{3cm} | p{14cm} | l }
\hline
Identifier & Sent By \\
\hline
Serial number, X-Apple-I-SRL-NO & \nolinkurl{gsa.apple.com/grandslam/MidService/startMachineProvisioning}, \nolinkurl{gsa.apple.com/grandslam/GsService2}, \nolinkurl{humb.apple.com/humbug/baa}, \nolinkurl{idiagnostics.apple.com/ios/v1.4/session} \\
\hline
IMEI  & \nolinkurl{play.itunes.apple.com/WebObjects/MZPlay.woa/wa/fuseNoCarrierBundleSubscription}, \nolinkurl{mask-api.icloud.com/v1/fetchAuthTokens}, \nolinkurl{albert.apple.com/deviceservices/activity/activation}, \nolinkurl{albert.apple.com/deviceservices/activity/phoneNumberSimNotification}, \nolinkurl{gsas.apple.com/grandslam/GsService2/postdata}\\
\hline 
EID, x-apple-seid & \nolinkurl{smp-device-content.apple.com/static/region/v2/config.json}, \nolinkurl{mask-api.icloud.com/v1/fetchAuthTokens}, \nolinkurl{p52-buy.itunes.apple.com/WebObjects/MZBuy.woa/wa/buyProduct} \\
\hline
X-Mme-Device-Id, guid, UDID, x-jmet-deviceid, baa-message & \nolinkurl{gsa.apple.com/grandslam/GsService2}, \nolinkurl{gsa.apple.com/grandslam/MidService/startMachineProvisioning}, \nolinkurl{gsa.apple.com/grandslam/MidService/finishMachineProvisioning}, \nolinkurl{humb.apple.com/humbug/baa}, \nolinkurl{acsegateway.icloud.com/v3/devices}, \nolinkurl{auth.itunes.apple.com/auth/v1/otp/1.5/finish}, \nolinkurl{musicstatus.itunes.apple.com/v1/status}, \nolinkurl{play.itunes.apple.com/WebObjects/MZPlay.woa/wa/fuseNoCarrierBundleSubscription}, \nolinkurl{p52-buy.itunes.apple.com/commerce/device/addPushNotificationType}, \nolinkurl{auth.itunes.apple.com/auth/v1/native}\\
\hline
X-Apple-I-MD-M & \\
\hline
X-Apple-I-AMD-M & \\
\hline
x-apple-baa-certs & \nolinkurl{kt-prod.ess.apple.com/kt_client/consistency_proof} \\
\hline
baa-certs & \nolinkurl{acsegateway.icloud.com/v3/devices}\\
\hline 
afds & \nolinkurl{p52-buy.itunes.apple.com/WebObjects/MZBuy.woa/wa/buyProduct} \\
\hline 
X-Apple-Baa & \nolinkurl{gsa.apple.com/grandslam/MidService/startMachineProvisioning}, \nolinkurl{gsa.apple.com/grandslam/MidService/finishMachineProvisioning}\\
\hline
\end{tabular}
\caption{Summary of device identifiers and the network connections in which we observe them being transmitted.}\label{tab:device_ids}
\end{table*}

\subsection{Device Identifiers}
In many network connections we observe various device identifiers being transmitted.  We briefly summarise these here.

The \vv{X-Apple-I-SRL-NO} value is the hardware serial number. The \vv{IMEI} uniquely identifies the SIM slot on the handset, and the \vv{EID} uniquely identifies the eSIM slot.  The \vv{UDID} value is the Apple Unique Device Identifier, an SHA1 hash of the serial number, ECID, Wifi and Bluetooth MAC addresses\footnote{\nolinkurl{https://en.wikipedia.org/wiki/UDID}}.  

The \vv{X-Apple-I-MD-M} value is the Anisette Machine Identifier\footnote{The terms Anisette and Machine Identifier come from analysis of the symbols within the Apple AuthKit private framework. See also \nolinkurl{https://github.com/ionescu007/Blackwood-4NT/blob/main/README.md}/}.  This is almost certainly a unique device identifier.  Similarly the \vv{X-Apple-I-AMD-M} value. 

%
%

The \vv{baa-certs} and \vv{x-apple-baa-certs} values are a base64 encoded Basic Attestation certificate that contains OID 1.2.840.113635.100.10.1 with the handset CHIP, IMEI, ECID, \vv{X-Apple-I-SRL-NO}, \vv{X-Mme-Device-Id}, WiFi and bluetooth MAC addresses.  Similarly, the \vv{afds} is a hex encoded Basic Attestation certificate.   It is not clear how the content of the \vv{X-Apple-Baa} value is calculated, likely it is encrypted.

\subsection{More Detail On Tracking of User Interactions}

\subsubsection{Tracking of User Searches}

Figure \ref{fig:xp_search} shows in more detail the data transmitted by the Apple App Store to \nolinkurl{xp.apple.com} when the user presses the ``Search'' button after typing``skin cancer app'' as a search query.   Observe that the search text `skin cancer app'' is transmitted as well as user identifiers including the \vv{DSID}.  Note that there is a separate connection to \nolinkurl{amp-api-search-edge.apps.apple.com/v1/catalog/ie/search/suggestions} which returns the search results. 

\begin{figure}[htb]
\centering
\begin{lstlisting}[frame=single,language=http]
POST https://xp.apple.com/report/2/xp_amp_appstore
!{\color{blue}Request Headers:}!
cookie : vrep=CIP...RAA, wosid-lite=6Eg...Jdg, xt-b-ts-!{\color{red}19129383683}!=1771423226373, pldfltcid=f5f...052, mz_at0-!{\color{red}19129383683}!=AwQ...i9Y=, fsas=AAA...u/tA==, xp_ab=1#KV...Ys01, itspod=52, mz_at_ssl-!{\color{red}19129383683}!=AwU...V90=, tv-pldfltcid=f5f5...052, X-Dsid=!{\color{red}19129383683}!, xt-src=b, xp_aci=!{\color{red}rgUruH9u2L0cz5BGpmyAqSEiAQ}!
xx-apple-i-md : AAA...AAw==
user-agent : AppStore/3.0 iOS/16.7.11 model/iPhone10,1 hwp/t8015 build/20H360 (6; dt:157) AMS/1
x-dsid : !{\color{red}19129383683}!
x-apple-client-application : com.apple.AppStore
x-apple-i-md-rinfo : 50660608
x-apple-i-md-m : !{\color{red}iiY...42L}!
x-apple-adsid :  !{\color{red}000191-10-d0cd9af9-d7b6-4b63-ac3b-71dd4e81c73f}!
!{\color{blue}POST Body:}!
{"postTime":1771423696267,"events":[{"anonymous":false,
<...>
"clientEventId":"!{\color{red}4F753CD5-FB64-4503-9979-9A2363A136C7}!","userNs":"appstore_clickstream_eval", "hardwareFamily":"iPhone", "actionType":"!{\color{green}submit}!", "xpUserIdSyncState":"unsynchronized", "eventTime":1771423689853, "hardwareModel":"iPhone10,1", "dsId":"!{\color{red}19129383683}!", "timezoneOffset":0, "appVersion":"3.0", "tabBadges":{"arcade":"1"}, "storeFrontHeader":"143449-2,29 t:apps3", "term":"!{\color{green}skin cancer app}!", "xpAccountsMatch":"false", "app":"com.apple.AppStore", "eventType":"!{\color{green}search}!", "osVersion":"16.7.11", "clientBuildType":"production", "eventVersion":3, "targetType":"!{\color{green}button}!", "topic":"xp_amp_appstore", "resourceRevNum":"22.4.4-3", "os":"iOS","osBuildNumber":"20H360" ,"connection":"WiFi", "clientId":"!{\color{red}DtWyjFCyDzlDBxHbfMEfJZHe}!", "pageContext":"Search", "baseVersion":1, "language":"2", "xpSendMethod":"jet-js", "userId":"!{\color{red}Aq3wN8EnjYCnET7jNsD52ssF}!"}
<...>
\end{lstlisting}
\caption{Data transmitted to \nolinkurl{xp.apple.com} when the user searches for ``skin cancer app'' in the Apple App Store. }\label{fig:xp_search}
\end{figure}

\subsubsection{Tracking of User Screen Views}

Figure \ref{fig:xp_view} shows in more detail the data transmitted by the Apple App Store to \nolinkurl{xp.apple.com} when the user views the search results after searching for ``skin cancer app'' in the Apple App Store.  In addition to a ``page'' event recording viewing of the search results, a separate ``impressions'' event is also transmitted recording the display of an advert.  Observe that advert identifiers transmitted include the \vv{impressionId}, \vv{metadata} and \vv{adamId} (see Section \ref{sec:search_ads} for a more detailed discussion of these identifiers) as well as user identifiers including the \vv{DSID}.

\begin{figure}[htb]
\centering
\begin{lstlisting}[frame=single,language=http]
POST https://xp.apple.com/report/2/xp_amp_appstore
!{\color{blue}Request Headers:}!
cookie : vrep=CIP...RAA, wosid-lite=6Eg...Jdg, xt-b-ts-!{\color{red}19129383683}!=1771423226373, pldfltcid=f5f...052, mz_at0-!{\color{red}19129383683}!=AwQ...i9Y=, fsas=AAA...u/tA==, xp_ab=1#KV...Ys01, itspod=52, mz_at_ssl-!{\color{red}19129383683}!=AwU...V90=, tv-pldfltcid=f5f5...052, X-Dsid=!{\color{red}19129383683}!, xt-src=b, xp_aci=!{\color{red}rgUruH9u2L0cz5BGpmyAqSEiAQ}!
xx-apple-i-md : AAA...AAw==
user-agent : AppStore/3.0 iOS/16.7.11 model/iPhone10,1 hwp/t8015 build/20H360 (6; dt:157) AMS/1
x-dsid : !{\color{red}19129383683}!
x-apple-client-application : com.apple.AppStore
x-apple-i-md-rinfo : 50660608
x-apple-i-md-m : !{\color{red}iiY...42L}!
x-apple-adsid :  !{\color{red}000191-10-d0cd9af9-d7b6-4b63-ac3b-71dd4e81c73f}!
!{\color{blue}POST Body:}!
{"postTime":1771423696267,"events":[{"anonymous":false,<...>
<...>"pageId":"!{\color{red}669...c4d}!","topic":"xp_amp_appstore","iAdId":"!{\color{red}AC515964-3260-447D-AA21-C86A3876135F}!","clientBuildType":"production","clientCorrelationKey":"3OE...V54","pageType":"Search","connection":"WiFi","xpUserIdSyncState":"unsynchronized","eventTime":1771423691237,"iAdAppStoreClientRequestId":"FCB03575-BE0A-4568-9426-63C66D5FDE06","clientId":"!{\color{red}DtWyjFCyDzlDBxHbfMEfJZHe}!","eventType":"!{\color{green}page}!", "pageContext":"Search","anonymous":false,"page":"Search_!{\color{red}669...c4d}!","searchTerm":"!{\color{green}skin cancer app}!", "userId":"!{\color{red}Aq3wN8EnjYCnET7jNsD52ssF}!","dsId":"!{\color{red}19129383683}!","iAdOdmlSuccess":false,"pageHistory":["Software_924418916","Software_924418916","Software_924418916","Search_fc31...62af","SearchLanding_SearchLanding"],"resourceRevNum":"22.4.4-3","iAdPlacementId":"!{\color{green}APPSTORE\_SEARCH\_RESULT\_PAGE}!","iAd":{"iAdFormat":{"userRating":true,"icon":"regular","images":true,"text":"promotionalText"},"iAdAlgoId":"0","iAdImpressionId":"!{\color{red}9d528845-f08d-3d26-90b0-893f673c35ef}!","iAdMetadata":"!{\color{red}CiQ...2Mg}!=="},"iAdRotationCount":0,"data.search.dataSetId":"!{\color{red}669...c4d}!",<...>"pageDetails":"Apps","adamId":"!{\color{red}1478978663}!","clientEventId":"71C...E38",<...>"pageUrl":"https:\/\/amp-api-search-edge.apps.apple.com\/v1\/catalog\/ie\/search?platform=iphone<...>&term=!{\color{green}skin cancer%20app}!<...>
<...>
\end{lstlisting}
\caption{Data transmitted to \nolinkurl{xp.apple.com} when the user views the result page for search ``skin cancer apps''.  The first search result is an ad, details of which are also transmitted.  In addition to this ``page'' event, a separate ``impressions'' event is transmitted recording the display of the advert.}\label{fig:xp_view}
\end{figure}

\subsubsection{Tracking of User Clicks}

Figure \ref{fig:xp_click} shows in more detail the data transmitted by the Apple App Store to \nolinkurl{xp.apple.com} when the user clicks the "Open" button in Apple App Store for the Scanoma skin cancer app.  Observe that advert identifiers transmitted include the \vv{impressionId}, \vv{metadata}, \vv{ppid}/\vv{pageCustomId} and \vv{adamId} as well as user identifiers including the \vv{DSID}.

\begin{figure}[htb]
\centering
\begin{lstlisting}[frame=single,language=http]
POST https://xp.apple.com/report/2/xp_amp_appstore
!{\color{blue}Request Headers:}!
cookie : mz_at0-!{\color{red}19129383683}!=AwQ...ORM=, wosid-lite=6Eg...Jdg, xt-b-ts-!{\color{red}19129383683}!=1771424149539, pldfltcid=f5f...052, tv-pldfltcid=f5...052, fsas=AAA...hu/tA==, xp_ab=1#KVpIi...Ys01, itspod=52, mz_at_ssl-!{\color{red}19129383683}!=AwU...G98=, X-Dsid=!{\color{red}19129383683}!, vrep=CIP...hAA, xt-src=b, xp_aci=!{\color{red}rgUruH9u2L0cz5BGpmyAqSEiAQ}!
x-apple-i-md : AAA...AAw==
user-agent : AppStore/3.0 iOS/16.7.11 model/iPhone10,1 hwp/t8015 build/20H360 (6; dt:157) AMS/1
x-dsid : !{\color{red}19129383683}!
x-apple-client-application : com.apple.AppStore
x-apple-i-md-rinfo : 50660608
x-apple-i-md-m : !{\color{red}iiY...42L}!
x-apple-adsid : !{\color{red} 000191-10-d0cd9af9-d7b6-4b63-ac3b-71dd4e81c73f}!
!{\color{blue}POST Body:}!
{"postTime":1771424209220,"events":[{"storeFront":"143449","osVersion":"16.7.11","pageId":"!{\color{red}1478978663}!","topic":"xp_amp_appstore","iAdId":"!{\color{red}AC515964-3260-447D-AA21-C86A3876135F}!","clientBuildType":"production","pageType":"Software","connection":"WiFi","clientId":"!{\color{red}DtWyjFCyDzlDBxHbfMEfJZHe}!","xpUserIdSyncState":"unsynchronized","eventTime":1771424209078,"iAdMetadata":"!{\color{red}CiQ...2Mg}!==","baseVersion":1,"eventType":"!{\color{red}click}!","xpAccountsMatch":"false","platformDisplayStyle":"ios","pageContext":"Search","anonymous":false,"hasiAdData":true,"hardwareFamily":"iPhone","hardwareModel":"iPhone10,1","iAdImpressionIndex":0,"page":"Software_1478978663","location":[{"id":"!{\color{red}1478978663}!","locationPosition":0,"locationType":"lockup","name":"GET","idType":"its_id"}],"app":"com.apple.AppStore","os":"iOS","userId":"!{\color{red}Aq3wN8EnjYCnET7jNsD52ssF}!","targetType":"!{\color{green}button}!","timezoneOffset":0,"dsId":"!{\color{red}19129383683}!","pageCustomId":"!{\color{red}5d6e3084-e2f3-42df-8224-3432a1b6d853}!","actionType":"!{\color{green}open}!","resourceRevNum":"22.4.4-3","iAdPlacementId":"APPSTORE_SEARCH_RESULT_PAGE","iAdImpressionId":"!{\color{red}9d528845-f08d-3d26-90b0-893f673c35ef}!","iAdRotationCount":0,"targetId":"!{\color{red}1478978663}!","xpSendMethod":"jet-js","eventVersion":4,"appVersion":"3.0","userNs":"appstore_clickstream_eval","iAdTemplateType":"PI3","pageDetails":"Scanoma Pty Ltd_scanoma - mole check","clientEventId":"!{\color{red}DD4F5209-4505-478F-BF4A-237A4A35AF2B}!","osBuildNumber":"20H360","storeFrontHeader":"143449-2,29 t:apps3","pageUrl":"https:\/\/apps.apple.com\/ie\/app\/scanoma-mole-check\/id1478978663?X-AppStore-iAdClickToken=%7B%22iAdId%22%3A%22!{\color{red}AC515964-3260-447D-AA21-C86A3876135F}!%22%2C%22iAdPlacementId%22%3A%22APPSTORE_SEARCH_RESULT_PAGE%22%2C%22hasiAdData%22%3Atrue%2C%22iAdImpressionId%22%3A%22!{\color{red}9d528845-f08d-3d26-90b0-893f673c35ef}!%22%2C%22iAdMetadata%22%3A%22!{\color{red}CiQ...J2Mg}!%3D%3D%22%2C%22iAdPageCustomId%22%3A%22!{\color{red}5d6e3084-e2f3-42df-8224-3432a1b6d853}!%22%2C%22iAdTemplateType%22%3A%22PI3%22%7D&ppid=!{\color{red}5d6e3084-e2f3-42df-8224-3432a1b6d853}!","language":"2","tabBadges":{"arcade":"1"},"subscriptionState":"unsubscribed","actionDetails":{"buyParams":"productType=C&price=0&salableAdamId=!{\color{red}1478978663}!&pricingParameters=STDQ&pg=default&appExtVrsId=!{\color{red}881173852}!&mtIadPlacementId=APPSTORE_SEARCH_RESULT_PAGE&mtIadTemplateType=PI3"},"iAdPageCustomId":"!{\color{red}5d6e3084-e2f3-42df-8224-3432a1b6d853}!"}],"deliveryVersion":1}
\end{lstlisting}
\caption{Data transmitted to \nolinkurl{xp.apple.com} when the user clicks the "Open" button in Apple App Store for the Scanoma skin cancer app.}\label{fig:xp_click}
\end{figure}

\end{document}

%% file: listings.tex
\usepackage{listings}

\lstdefinelanguage{logsources}{
keywords=[1]{CARRIER_SERVICES, ANDROID_DIALER,SCOOBY_EVENT,ANDROID_MESSAGING}
}

\lstdefinelanguage{pseudo}{
    keywords=[1]{androidID,pseudonymousIdToken,NIDCookie,logSourceName,incomingPhoneNumber},
    keywords=[2]{header,deviceStatus,deviceInfo,dynamiteModules,accessibilitySettings,googlePlayServicesVersionName,sdkExtensions,logEntry,event},
    keywords=[3]{SDKVersion,currentTimeMillis,qosTier,elapsedTime,model,product,buildID,hardware,device,simOperator,language,country,manufacturer,brand,board,radioVersion,buildFingerprint,buildType,simCarrierId,timestamp,networkConnected,batteryCharged,accountIsGoogleCom,googlePlayServicesVersionString,googlePlayServicesVersionCode,packageVersionName,packageName},
}
\lstdefinelanguage{pseudo2}{
    keywords=[1]{AOSPEventType,timestamp,callDuration,callID},
    keywords=[2]{logEntry,event,experimentTokens,tokens,connectionDetail,deviceDetails,impressionEvent,connectionDetails,searchQuery,callDetails,callSetupTiming},
    keywords=[3]{buildDevice,buildModel,buildID,systemApp,deviceClass,token,tokenBytes,hasContent,bootCount,elapsedRealtime,subEvent,tz_offset,buildVersionRelease,releaseStatus,activeNetworkType,updatedSystemApp,simOperator,mobileOperator,country,installedBy,callInitiationType,isIncoming,lookupResultType,disconnectCause,queryLength,searchTimeMillis,elapsedDaysSinceDialerInstall},
}
\lstdefinelanguage{pseudo3}{
    keywords=[1]{eventType,timestamp,sha256HashMsg,sha256HashPrevMsg,currentTime_ms,elapsedTimeSinceMsgSendRecv_ms},
    keywords=[2]{logEntry,event,experimentTokens,tokens,bugleMessage,messageProtocol,bugleMessageStatus,subscriptionUtils,mimeData,messageTiming,phoneNumberFormat2,carrierServicesData,messageSettings,mimeData2,connectionDetails},
    keywords=[3]{subEvent,tz_offset,bootCount,elapsedRealtime,activeNetworkType,messageSendClickToSentLatency,bugleMessageSource,usageStatsLoggingId,conversationType,sendAttempt,wasRCS,rcsStatus,configStatus,elapsedTimeEnum,RcsConfig,messageStatus,conversationIdSHA1,Mcc,Mnc,simOperatorName,networkOperatorName,simCarrierId,mimeType,valid,messagePart,versionCode,hasParticpants,hasConversationId,sendModeIsAuto,sendModeIsXms,onlineStatus,sendmode,token,tokenBytes,hasContent}
}

\lstdefinelanguage{http}{
    alsoletter=-,
    keywords=[1]{DSID},
    keywordstyle=[2]\color{blue}\bfseries,
    keywords=[2]{},
    keywordstyle=[3]\color{brown},
    keywords=[3]{authorization, Authorization, user-agent, User-Agent, x-server-token, set-cookie, device:, x-goog-spatula, x-ps-rh, X-Android-Package},
    keywordstyle=[4]\color{blue}\bfseries,
    keywords=[4]{POST, GET},
}
\lstdefinelanguage{dialer}{
   basicstyle=\tiny\tt\color{brown},
    alsoletter=_,
    keywords=[1]{},
    keywords=[3]{callDuration},
    keywords=[2]{MAIN_CLICK_FAB_TO_OPEN_DIALPAD,queryLength,PRECALL_INITIATED,DIALER_GATEWAY_CALL_INFO_LOOKUP_SERVICE_PHONE_LOOKUP_UNAUTHENTICATED,DIALER_GATEWAY_CALL_INFO_LOOKUP_SERVICE_PHONE_LOOKUP_FAILED,TIDEPODS_STATUS_BAR_NOTIFICATION_SHOWED,TIDEPODS_BUBBLE_SHOWED,USER_PARTICIPATED_IN_A_CALL,ANNOTATED_CALL_LOG_FORCE_REFRESH_CHANGES_NEEDED,ANNOTATED_CALL_LOG_FORCE_REFRESH_NO_CHANGES_NEEDED, ANNOTATED_CALL_LOG_NOT_DIRTY,CALL_SCREENING_SERVICE_MUSIC_IS_NOT_ACTIVE,INCOMING_CALL_SCREENED,INCOMING_NON_SPAM_CALL,DIALER_GATEWAY_CALL_INFO_LOOKUP_SERVICE_PHONE_LOOKUP_FAILED,DARK_MODE_SET_FOLLOW_SYSTEM_BY_CONTACTS,IN_CALL_DIALPAD_HANG_UP_BUTTON_PRESSED,NON_SPAM_NOTIFICATION_NOT_SHOWN_AFTER_THROTTLE},
}
\lstdefinelanguage{messages}{
   basicstyle=\tiny\tt\color{brown},
    alsoletter=_,
    keywords=[3]{sendAttempt,conversationIdSHA1},
    keywords=[1]{sha256HashMsg,sha256HashPrevMsg,simSerialNumber},
    keywords=[2]{BUGLE_MESSAGE,BUGLE_APP_CONFIGURATION,BUGLE_COMPOSE,BUGLE_P2P_SUGGESTION,BUGLE_SPAM_DETECTION,BUGLE_TELEPHONY_EVENT,BUGLE_APP,BUGLE_CONVERSATION,BUGLE_CONTACT_BANNER},
}
\lstdefinelanguage{analytics}{
   basicstyle=\tiny\tt\color{brown},
    alsoletter=-,
    keywords=[1]{event_timestamp,google_ad_id,firebase_instance_id,package_name},
    keywords=[2]{bundle,event_info,user_info,event},
    keywords=[3]{message_timestamp,bundle_end_timestamp,last_bundle_end_timestamp,operating_system,operating_system_version,Build_MODEL,language_country,timezone_offset_mins,app_version,gmp_version,gms_version,random_hex,dev_cert_hash,daily_conversions_count,gmp_app_id,last_bundle_start_timestamp2,firebase_instance_id,app_version_int,config_version,M,dynamite_version},
}
\lstdefinelanguage{analytics_event}{
   basicstyle=\tiny\tt\color{brown},
    alsoletter=-,
    keywords=[1]{event_timestamp,event_code},
    keywords=[2]{event, event_info},
    keywords=[3]{setting_code,data_str,previous_event_timestamp},
}
\lstdefinelanguage{analytics_dialer}{
   basicstyle=\tiny\tt\color{brown},
    alsoletter=-,
    keywords=[3]{event_code,data_str},
}

%% file: appleappstore.bbl
\begin{thebibliography}{10}
\providecommand{\url}[1]{#1}
\csname url@samestyle\endcsname
\providecommand{\newblock}{\relax}
\providecommand{\bibinfo}[2]{#2}
\providecommand{\BIBentrySTDinterwordspacing}{\spaceskip=0pt\relax}
\providecommand{\BIBentryALTinterwordstretchfactor}{4}
\providecommand{\BIBentryALTinterwordspacing}{\spaceskip=\fontdimen2\font plus
\BIBentryALTinterwordstretchfactor\fontdimen3\font minus
  \fontdimen4\font\relax}
\providecommand{\BIBforeignlanguage}[2]{{%
\expandafter\ifx\csname l@#1\endcsname\relax
\typeout{** WARNING: IEEEtran.bst: No hyphenation pattern has been}%
\typeout{** loaded for the language `#1'. Using the pattern for}%
\typeout{** the default language instead.}%
\else
\language=\csname l@#1\endcsname
\fi
#2}}
\providecommand{\BIBdecl}{\relax}
\BIBdecl

\bibitem{edpbcookies2012}
\BIBentryALTinterwordspacing
``{Opinion 04/2012 on Cookie Consent Exemption, Article 29 Data Protection
  Working Party},'' June 2012. [Online]. Available:
  \url{https://ec.europa.eu/justice/article-29/documentation/opinion-recommendation/files/2012/wp194_en.pdf}
\BIBentrySTDinterwordspacing

\bibitem{gdpr}
\BIBentryALTinterwordspacing
``{European Parliament and Council. Regulation 2016/679 (General Data
  Protection Regulation)},'' 2016. [Online]. Available:
  \url{https://data.europa.eu/eli/reg/2016/679/oj}
\BIBentrySTDinterwordspacing

\bibitem{edpbguidelines}
\BIBentryALTinterwordspacing
``{Guidelines 2/2023 on Technical Scope of Art. 5(3) of ePrivacy Directive,
  EDPB},'' Oct 2024. [Online]. Available:
  \url{https://www.edpb.europa.eu/system/files/2024-10/edpb_guidelines_202302_technical_scope_art_53_eprivacydirective_v2_en_0.pdf}
\BIBentrySTDinterwordspacing

\bibitem{edpbinterplay}
\BIBentryALTinterwordspacing
``{Opinion 5/2019 on the interplay between the ePrivacy Directive and the GDPR,
  in particular regarding the competence, tasks and powers of data protection
  authorities. EDPB.}'' March 2019. [Online]. Available:
  \url{https://www.edpb.europa.eu/sites/default/files/files/file1/201905_edpb_opinion_eprivacydir_gdpr_interplay_en_0.pdf}
\BIBentrySTDinterwordspacing

\bibitem{edpbconsent}
\BIBentryALTinterwordspacing
``{Guidelines 05/2020 on consent under Regulation 2016/679, EDPB},'' May 2020.
  [Online]. Available:
  \url{https://www.edpb.europa.eu/sites/default/files/files/file1/edpb_guidelines_202005_consent_en.pdf}
\BIBentrySTDinterwordspacing

\bibitem{irishlaw}
\BIBentryALTinterwordspacing
``{Report by the Data Protection Commission on the use of cookies and other
  tracking technologies},'' April 2020. [Online]. Available:
  \url{https://www.irishstatutebook.ie/eli/2011/si/336/}
\BIBentrySTDinterwordspacing

\bibitem{edpbcookiebanners2023}
\BIBentryALTinterwordspacing
``{Report of the work undertaken by the Cookie Banner Taskforce, EDPB},'' Jan
  2023. [Online]. Available:
  \url{https://www.edpb.europa.eu/system/files/2023-01/edpb_20230118_report_cookie_banner_taskforce_en.pdf}
\BIBentrySTDinterwordspacing

\bibitem{gdpr_eu}
\BIBentryALTinterwordspacing
``{Legal grounds for processing data, European Commission}.'' [Online].
  Available:
  \url{https://commission.europa.eu/law/law-topic/data-protection/rules-business-and-organisations/legal-grounds-processing-data_en}
\BIBentrySTDinterwordspacing

\bibitem{edpblegitimateinterests}
\BIBentryALTinterwordspacing
``{Guidelines 1/2024 on processing of personal data based on Article 6(1)(f)
  GDPR, EDPB},'' Oct 2024. [Online]. Available:
  \url{https://www.edpb.europa.eu/system/files/2024-10/edpb_guidelines_202401_legitimateinterest_en.pdf}
\BIBentrySTDinterwordspacing

\bibitem{dpcguidance}
\BIBentryALTinterwordspacing
``{Guidance Notes: Legal Bases for Processing Personal Data, Irish Data
  Protection Commision},'' Dec 2019. [Online]. Available:
  \url{https://www.dataprotection.ie/sites/default/files/uploads/2020-04/Guidance%20on%20Legal%20Bases.pdf}
\BIBentrySTDinterwordspacing

\bibitem{mitmproxy}
\BIBentryALTinterwordspacing
A.~Cortesi, M.~Hils, T.~Kriechbaumer, and contributors, ``{mitmproxy}: A free
  and open source interactive {HTTPS} proxy (v5.01),'' 2020. [Online].
  Available: \url{https://mitmproxy.org/}
\BIBentrySTDinterwordspacing

\bibitem{dpc_faq}
\BIBentryALTinterwordspacing
``{FAQ on Consent for Electronic Direct Marketing, Irish Data Protection
  Commission},'' April 2020. [Online]. Available:
  \url{https://dataprotection.ie/sites/default/files/uploads/2020-05/FAQ on
  Consen for Electronic Direct Marketing - April2020.pdf}
\BIBentrySTDinterwordspacing

\bibitem{10.1145/2872427.2882991}
A.~Cahn, S.~Alfeld, P.~Barford, and S.~Muthukrishnan, ``{An Empirical Study of
  Web Cookies},'' in \emph{Proc WWW}, 2016, pp. 891--901.

\bibitem{10.1145/2736277.2741679}
\BIBentryALTinterwordspacing
S.~Englehardt, D.~Reisman, C.~Eubank, P.~Zimmerman, J.~Mayer, A.~Narayanan, and
  E.~W. Felten, ``Cookies that give you away: The surveillance implications of
  web tracking,'' in \emph{Proceedings of the 24th International Conference on
  World Wide Web}, ser. WWW '15.\hskip 1em plus 0.5em minus 0.4em\relax
  Republic and Canton of Geneva, CHE: International World Wide Web Conferences
  Steering Committee, 2015, pp. 289--299. [Online]. Available:
  \url{https://doi.org/10.1145/2736277.2741679}
\BIBentrySTDinterwordspacing

\bibitem{10.1145/3176246}
\BIBentryALTinterwordspacing
R.~Binns, J.~Zhao, M.~V. Kleek, and N.~Shadbolt, ``Measuring third-party
  tracker power across web and mobile,'' \emph{ACM Trans. Internet Technol.},
  vol.~18, no.~4, Aug. 2018. [Online]. Available:
  \url{https://doi.org/10.1145/3176246}
\BIBentrySTDinterwordspacing

\bibitem{203894}
\BIBentryALTinterwordspacing
S.~Zimmeck, J.~S. Li, H.~Kim, S.~M. Bellovin, and T.~Jebara, ``A privacy
  analysis of cross-device tracking,'' in \emph{26th USENIX Security Symposium
  (USENIX Security 17)}.\hskip 1em plus 0.5em minus 0.4em\relax Vancouver, BC:
  USENIX Association, Aug. 2017, pp. 1391--1408. [Online]. Available:
  \url{https://www.usenix.org/conference/usenixsecurity17/technical-sessions/presentation/zimmeck}
\BIBentrySTDinterwordspacing

\bibitem{OmniCrawl}
D.~Cassel, S.-C. Lin, A.~Buraggina, W.~Wang, A.~Zhang, L.~Bauer, H.-C. Hsiao,
  and L.~J.~T. Libert, ``Omnicrawl: Comprehensive measurement of web tracking
  with real desktop and mobile browsers,'' in \emph{Proc PETS}, 2022, pp.
  227--252.

\bibitem{8002896}
R.~Gonzalez, L.~Jiang, M.~Ahmed, M.~Marciel, R.~Cuevas, H.~Metwalley, and
  S.~Niccolini, ``{The cookie recipe: Untangling the use of cookies in the
  wild},'' in \emph{Proc TMA}, 2017, pp. 1--9.

\bibitem{10.1145/3501247.3531545}
M.~Gotze, S.~Matic, C.~Iordanou, G.~Smaragdakis, and N.~Laoutaris, ``{Measuring
  Web Cookies in Governmental Websites},'' in \emph{Proc 4th ACM Web Science
  Conference}, ser. WebSci '22, 2022, pp. 44--54.

\bibitem{10.1145/3442381.3449837}
\BIBentryALTinterwordspacing
Q.~Chen, P.~Ilia, M.~Polychronakis, and A.~Kapravelos, ``Cookie swap party:
  Abusing first-party cookies for web tracking,'' in \emph{Proceedings of the
  Web Conference 2021}, ser. WWW '21.\hskip 1em plus 0.5em minus 0.4em\relax
  New York, NY, USA: Association for Computing Machinery, 2021, pp. 2117--2129.
  [Online]. Available: \url{https://doi.org/10.1145/3442381.3449837}
\BIBentrySTDinterwordspacing

\bibitem{8706882}
G.~Franken, T.~Van~Goethem, and W.~Joosen, ``Exposing cookie policy flaws
  through an extensive evaluation of browsers and their extensions,''
  \emph{IEEE Security \& Privacy}, vol.~17, no.~4, pp. 25--34, 2019.

\bibitem{10.1145/3386040}
\BIBentryALTinterwordspacing
P.~Laperdrix, N.~Bielova, B.~Baudry, and G.~Avoine, ``Browser fingerprinting: A
  survey,'' \emph{ACM Trans. Web}, vol.~14, no.~2, Apr. 2020. [Online].
  Available: \url{https://doi.org/10.1145/3386040}
\BIBentrySTDinterwordspacing

\bibitem{10.1145/3576915.3616586}
\BIBentryALTinterwordspacing
S.~Munir, S.~Siby, U.~Iqbal, S.~Englehardt, Z.~Shafiq, and C.~Troncoso,
  ``Cookiegraph: Understanding and detecting first-party tracking cookies,'' in
  \emph{Proceedings of the 2023 ACM SIGSAC Conference on Computer and
  Communications Security}, ser. CCS '23.\hskip 1em plus 0.5em minus
  0.4em\relax New York, NY, USA: Association for Computing Machinery, 2023, pp.
  3490--3504. [Online]. Available:
  \url{https://doi.org/10.1145/3576915.3616586}
\BIBentrySTDinterwordspacing

\bibitem{7872467}
T.~Bujlow, V.~Carela-Espanol, J.~Sole-Pareta, and P.~Barlet-Ros, ``A survey on
  web tracking: Mechanisms, implications, and defenses,'' \emph{Proc IEEE},
  vol. 105, no.~8, pp. 1476--1510, 2017.

\bibitem{217648}
\BIBentryALTinterwordspacing
G.~Franken, T.~V. Goethem, and W.~Joosen, ``Who left open the cookie jar? a
  comprehensive evaluation of {Third-Party} cookie policies,'' in \emph{27th
  USENIX Security Symposium (USENIX Security 18)}.\hskip 1em plus 0.5em minus
  0.4em\relax Baltimore, MD: USENIX Association, Aug. 2018, pp. 151--168.
  [Online]. Available:
  \url{https://www.usenix.org/conference/usenixsecurity18/presentation/franken}
\BIBentrySTDinterwordspacing

\bibitem{10.1007/978-3-030-15986-3_17}
A.~Dabrowski, G.~Merzdovnik, J.~Ullrich, G.~Sendera, and E.~Weippl, ``Measuring
  cookies and web privacy in a post-gdpr world,'' in \emph{Passive and Active
  Measurement}, D.~Choffnes and M.~Barcellos, Eds.\hskip 1em plus 0.5em minus
  0.4em\relax Cham: Springer International Publishing, 2019, pp. 258--270.

\bibitem{10.1145/3708515}
\BIBentryALTinterwordspacing
N.~Singh, Y.~Do, Y.~Yu, I.~Fouad, J.~Kim, and H.~Kim, ``Crumbled cookies:
  Exploring e-commerce websites cookie policies with data protection
  regulations,'' \emph{ACM Trans. Web}, vol.~19, no.~1, Jan. 2025. [Online].
  Available: \url{https://doi.org/10.1145/3708515}
\BIBentrySTDinterwordspacing

\bibitem{10.1145/3442381.3450056}
\BIBentryALTinterwordspacing
E.~Papadogiannakis, P.~Papadopoulos, N.~Kourtellis, and E.~P. Markatos, ``User
  tracking in the post-cookie era: How websites bypass gdpr consent to track
  users,'' in \emph{Proceedings of the Web Conference 2021}, ser. WWW
  '21.\hskip 1em plus 0.5em minus 0.4em\relax New York, NY, USA: Association
  for Computing Machinery, 2021, pp. 2130--2141. [Online]. Available:
  \url{https://doi.org/10.1145/3442381.3450056}
\BIBentrySTDinterwordspacing

\bibitem{4yearscookies}
M.~Trevisan, S.~Traverso, E.~Bassi, and M.~Mellia, ``{4 Years of EU Cookie Law:
  Results and Lessons Learned},'' in \emph{Proc Privacy Enhancing Technologies
  Symposium}, no.~2, 2019, pp. 126--145.

\bibitem{ndss2019}
M.~Degeling, C.~Utz, C.~Lentzsch, H.~Hosseini, F.~Schaub, and T.~Holz, ``{We
  Value Your Privacy ... Now Take Some Cookies: Measuring the GDPR's Impact on
  Web Privacy},'' in \emph{Proc NDSS}, 2019.

\bibitem{irishdpc}
\BIBentryALTinterwordspacing
``{S.I. No. 336/2011},'' 2011. [Online]. Available:
  \url{https://www.dataprotection.ie/sites/default/files/uploads/2020-04/Data%20Protection%20Commission%20cookies%20sweep%20REVISED%2015%20April%202020%20v.01.pdf}
\BIBentrySTDinterwordspacing

\bibitem{edpbcookies}
\BIBentryALTinterwordspacing
``{Report of the work undertaken by the Cookie Banner Taskforce, European Data
  Protection Board},'' Jan 2023. [Online]. Available:
  \url{https://www.edpb.europa.eu/system/files/2023-01/edpb_20230118_report_cookie_banner_taskforce_en.pdf}
\BIBentrySTDinterwordspacing

\bibitem{277190}
\BIBentryALTinterwordspacing
D.~Bollinger, K.~Kubicek, C.~Cotrini, and D.~Basin, ``Automating cookie consent
  and {GDPR} violation detection,'' in \emph{31st USENIX Security Symposium
  (USENIX Security 22)}.\hskip 1em plus 0.5em minus 0.4em\relax Boston, MA:
  USENIX Association, Aug. 2022, pp. 2893--2910. [Online]. Available:
  \url{https://www.usenix.org/conference/usenixsecurity22/presentation/bollinger}
\BIBentrySTDinterwordspacing

\bibitem{10.1007/978-3-031-28486-1_26}
A.~Rasaii, S.~Singh, D.~Gosain, and O.~Gasser, ``Exploring the cookieverse: A
  multi-perspective analysis of web cookies,'' in \emph{Passive and Active
  Measurement}, A.~Brunstrom, M.~Flores, and M.~Fiore, Eds.\hskip 1em plus
  0.5em minus 0.4em\relax Cham: Springer Nature Switzerland, 2023, pp.
  623--651.

\bibitem{274431}
\BIBentryALTinterwordspacing
K.~Kollnig, P.~Dewitte, M.~V. Kleek, G.~Wang, D.~Omeiza, H.~Webb, and
  N.~Shadbolt, ``A fait accompli? an empirical study into the absence of
  consent to {Third-Party} tracking in android apps,'' in \emph{Seventeenth
  Symposium on Usable Privacy and Security (SOUPS 2021)}.\hskip 1em plus 0.5em
  minus 0.4em\relax USENIX Association, Aug. 2021, pp. 181--196. [Online].
  Available:
  \url{https://www.usenix.org/conference/soups2021/presentation/kollnig}
\BIBentrySTDinterwordspacing

\bibitem{274614}
\BIBentryALTinterwordspacing
T.~T. Nguyen, M.~Backes, N.~Marnau, and B.~Stock, ``Share first, ask later (or
  never?) studying violations of {GDPR{\textquoteright}s} explicit consent in
  android apps,'' in \emph{30th USENIX Security Symposium (USENIX Security
  21)}.\hskip 1em plus 0.5em minus 0.4em\relax USENIX Association, Aug. 2021,
  pp. 3667--3684. [Online]. Available:
  \url{https://www.usenix.org/conference/usenixsecurity21/presentation/nguyen}
\BIBentrySTDinterwordspacing

\bibitem{iphones}
K.~Kollnig, A.~Shuba, R.~Binns, M.~V. Kleek, and N.~Shadbolt, ``Are iphones
  really better for privacy? a comparative study of ios and android apps,'' in
  \emph{Proc PETS}, 2022, pp. 6--24.

\bibitem{infocomgaen21}
D.~J. Leith and S.~Farrell, ``{Contact Tracing App Privacy: What Data Is Shared
  By Europe's GAEN Contact Tracing Apps},'' in \emph{Proc IEEE INFOCOM}, 2021.

\bibitem{securecom21}
D.~J. Leith, ``{Mobile Handset Privacy: Measuring The Data iOS and Android Send
  to Apple And Google},'' in \emph{Proc Securecomm}, 2021.

\bibitem{snooping21}
H.~Liu, P.~Patras, and D.~J. Leith, ``{ On the data privacy practices of
  Android OEMs},'' \emph{PLoS ONE}, 2023.

\bibitem{securecom22}
D.~J. Leith, ``{What Data Do The Google Dialer and Messages Apps On Android
  Send to Google?}'' in \emph{Proc Securecomm}, 2022.

\bibitem{narseo18}
A.~Razaghpanah, R.~Nithyanand, N.~Vallina-Rodriguez, and S.~Sundaresan,
  ``{Apps, Trackers, Privacy, and Regulators: A Global Study of the Mobile
  Tracking Ecosystem},'' in \emph{Proc NDSS}, 2018.

\bibitem{LEITH2026104850}
\BIBentryALTinterwordspacing
D.~Leith, ``Cookies, identifiers and other data that google silently stores on
  android handsets,'' \emph{Computers \& Security}, vol. 165, p. 104850, 2026.
  [Online]. Available:
  \url{https://www.sciencedirect.com/science/article/pii/S016740482600026X}
\BIBentrySTDinterwordspacing

\end{thebibliography}
